\documentclass[journal]{IEEEtran}
\IEEEoverridecommandlockouts
\usepackage{acronym}
\newacro{lstm}[LSTM]{long short term memory}
\newacro{has}[HAS]{HTTP adaptive streaming}
\newacro{hevc}[HEVC]{high-efficiency video coding}
\newacro{hm}[HM]{head movement}
\newacro{em}[EM]{eye movement}
\newacro{e2e}[E2E]{end-to-end}
\newacro{g2a}[G2A]{glass-to-algorithm}
\newacro{g2g}[G2G]{glass-to-glass}
\newacro{m2p}[M2P]{motion-to-photon}
\newacro{los}[LOS]{line of sight}

\newacro{ott}[OTT]{over-the-top}
\newacro{ml}[ML]{machine learning}
\newacro{dl}[DL]{deep learning}
\newacro{rq}[RQ]{rate-quality}
\newacro{rd}[RD]{rate-distortion}
\newacro{pf}[PF]{Pareto front}
\newacro{vod}[VOD]{video on demand}
\newacro{qp}[QP]{quantization parameter}
\newacro{qoe}[QoE]{quality of experience}
\newacro{bd-br}[BD-BR]{Bjøntegaard delta bitrate}
\newacro{hls}[HLS]{HTTP live streaming}
\newacro{dash}[DASH]{dynamic adaptive streaming over HTTP}
\newacro{vmaf}[VMAF]{video multi-method assessment fusion}
\newacro{rnn}[RNN]{recurrent neural network}
\newacro{crf}[CRF]{constant rate factor}
\newacro{vbr}[VBR]{variable bit-rate}
\newacro{abr}[ABR]{adaptive bit-rate}
\newacro{cbr}[CBR]{constant bit-rate}
\newacro{cqp}[CQP]{constant QP}
\newacro{jnd}[JND]{just noticeable difference}
\newacro{svr}[SVR]{support vector regression}
\newacro{gpr}[GPR]{Gaussian processes regression}
\newacro{rfr}[RFR]{random forest regression}
\newacro{si}[SI]{spatial information}
\newacro{ti}[TI]{temporal information}
\newacro{cf}[CF]{colorfulness}
\newacro{glcm}[GLCM]{gray level co-occurrence matrix }
\newacro{tc}[TC]{temporal coherence}
\newacro{ncc}[NCC]{normalized cross correlation}
\newacro{rfe}[RFE]{recursive feature elimination}
\newacro{gp}[GP]{Gaussian process}
\newacro{cnn}[CNN]{convolutional neural network}
\newacro{al}[AL]{Apple ladder}
\newacro{rl}[RL]{reference ladder}
\newacro{gt}[GT]{ground truth}
\newacro{uhd}[UHD]{ultra high definition}
\newacro{plcc}[PLCC]{Pearson linear correlation coefficient}
\newacro{srocc}[SROCC]{Spearman rank order correlation coefficient}
\newacro{vvc}[VVC]{versatile video coding}
\newacro{vvenc}[VVenC]{fraunhofer versatile video encoder}
\newacro{avc}[AVC]{advanced video coding}
\newacro{av1}[AV1]{AOMedia video 1}
\newacro{vp9}[VP9]{VP9}
\newacro{fpv}[FPV]{first person view}

\newacro{rapl}[RAPL]{running average power limit}
\newacro{soc}[SoC]{system-on-chip }
\newacro{ssim}[SSIM]{structural similarity index measure}
\newacro{psnr}[PSNR]{peak signal to noise ratio}
\newacro{vmaf}[VMAF]{video multi-method assessment fusion}
\newacro{mse}[MSE]{mean squared error}
\newacro{cpu}[CPU]{central processing unit}
\newacro{gpu}[GPU]{graphics processing unit}
\newacro{gcp}[GCP]{google cloud platform}
\newacro{IaaS}[IaaS]{infrastructure  as  a  service}
\newacro{fps}[fps]{frames per second}
\newacro{itu-t}[ITU-T]{international telecommunication Union - telecommunication standardization sector}
\newacro{mpeg}[MPEG]{moving picture experts group}
\newacro{hvs}[HVS]{human vision system }
\newacro{bd-rate}[BD-rate]{bjøntegaard delta rate}
\newacro{co2}[CO2]{carbon dioxide}
\newacro{mpeg}[MPEG]{motion picture experts group}
\newacro{vqeg}[VQEG]{video coding experts group}
\newacro{jct-vc}[JCT-VC]{joint collaborative team on video coding}
\newacro{aom}[AOM]{alliance for open media }
\newacro{ffmpeg}[FFmpeg]{fast forward moving picture experts group}
\newacro{jvet-ctc}[JVET-CTC]{joint video exploration team common test conditions}

\usepackage{cite}
\usepackage{amsmath,amssymb,amsfonts}
\usepackage{algorithmic}
\usepackage{graphicx}
\usepackage{textcomp}
\usepackage{soul}
\usepackage{makecell}
\usepackage{booktabs} 
\usepackage{multirow}
\usepackage{adjustbox}
\usepackage{subfigure, soul}

\newcommand{\tabitem}{~~\llap{\textbullet}~~}

\usepackage{array,booktabs}
\graphicspath{./figures/}
\usepackage[thinlines]{easytable}
\usepackage{bbding}
\usepackage{array,multirow,makecell}
\usepackage[table]{xcolor}

\definecolor{grannysmithapple}{rgb}{0.66, 0.89, 0.63}
\definecolor{green(colorwheel)(x11green)}{rgb}{0.0, 1.0, 0.0}
\definecolor{lightgreen}{rgb}{0.56, 0.93, 0.56}
\definecolor{lightcoral}{rgb}{0.94, 0.5, 0.5}

\definecolor{bl0}{rgb}{0.36, 0.54, 0.66}
\definecolor{bl1}{rgb}{0.94, 0.97, 1.0}

\graphicspath{ {./Figures/} }
\usepackage[thinlines]{easytable}
\usepackage{adjustbox}
\usepackage{acronym}
\def\BibTeX{{\rm B\kern-.05em{\sc i\kern-.025em b}\kern-.08em
    T\kern-.1667em\lower.7ex\hbox{E}\kern-.125emX}}
    
\usepackage{hyperref}

\usepackage{nccmath}
\usepackage[edges]{forest}
\usetikzlibrary{arrows.meta, shapes.geometric, calc, shadows, shadows.blur} 

\usepackage{array, etoolbox}
\usepackage{multido}

\newcommand{\blackbullets}[1]{\multido{\i=1+1}{3}{\ifnumgreater{\i}{#1}{\color{bl1}}{\color{bl0}}\Large \textbullet\kern 0.1em}}

\hypersetup{
    colorlinks=false,
    linkcolor=black,
    filecolor=black,      
    urlcolor=black,
}

\newacro{vvc}[VVC]{versatile video coding}
\newacro{hmd}[HMD]{head-mounted display}
\newacro{hevc}[HEVC]{high-efficiency video coding}
\newacro{odv}[ODV]{omnidirectional video}
\newacro{qoe}[QoE]{quality of experience}
\newacro{mv}[MV]{motion vector}
\newacro{mcts}[MCTS]{motion-constrained tile set}
\newacro{sei}[SEI]{supplemental enhancement information}
\newacro{sei}[SEI]{supplemental enhancement information}
\newacro{cmp}[CMP]{cube map projection}
\newacro{tsp}[TSP]{truncated square pyramid}
\newacro{erp}[ERP]{equirectangular projection}
\newacro{3dof}[3DoF]{three degrees of freedom}
\newacro{6dof}[6DoF]{six degrees of freedom}
\newacro{uav}[UAV]{unmanned aerial vehicle}
\newacro{6dof}[6DoF]{six degrees of freedom}
\newacro{srd}[SRD]{spatial relationship description}
\newacro{omaf}[OMAF]{omnidirectional media format}
\newacro{webrtc}[WebRTC]{web real-time communication}
\newacro{isobmff}[ISOBMFF]{ISO base media file format}
\newacro{dash}[DASH]{dynamic adaptive streaming over HTTP}
\newacro{rwp}[RWP]{region-wise packing}
\newacro{avc}[AVC]{advanced video coding}
\newacro{ann}[ANN]{artificial neural network}
\newacro{rwqr}[RWQR]{region-wise quality ranking}
\newacro{ivo}[IVO]{initial viewing orientation}
\newacro{rvtm}[RVTM]{recommended viewport timed metadata}
\newacro{vs}[VS]{viewport-specific}
\newacro{srtp}[SRTP]{secure real-time transport protocol}
\newacro{rtcp}[RTCP]{real-time transport control protocol}
\newacro{av1}[AV1]{AOMedia video 1}
\newacro{eeo}[EEO]{entropy equilibrium optimization}
\newacro{vr}[VR]{virtual reality}
\newacro{fov}[FoV]{field-of-view}
\newacro{eac}[EAC]{equi-angular cubemap}
\newacro{3gpp}[3GPP]{Third Generation Partnership Project}

\newacro{lte}[LTE]{Long Term Evolution}

\newacro{sw}[SW]{Software}
\newacro{hw}[HW]{Hardware}

\newacro{ai}[AI]{artificial intelligence}
\newacro{cnn}[CNN]{convolution neural network}
\newacro{mmwave}[mmWave]{millimeter wave}
\newacro{bs}[BS]{base station}
\newacro{psnr}[PSNR]{peak signal-to-noise rate}
\newacro{ssim}[SSIM]{structural similarity index measure}
\newacro{mimo}[MIMO]{multiple-input and multiple-output}
\newacro{e2e}[E2E]{end-to-end}
\newacro{g2g}[G2G]{glass-to-glass}
\newacro{2d}[2D]{two-dimensional}
\newacro{3d}[3D]{three-dimensional}
\newacro{svc}[SVC]{scalable video coding}
\newacro{ap}[AP]{access point}
\newacro{los}[LoS]{line-of-sight}
\newacro{nlos}[NLoS]{non-line-of-sight}
\newacro{rtp}[RTP]{real-time transport protocol}
\newacro{fpv}[FPV]{First Person View}
\newacro{svt-av1}[SVT-AV1]{scalable video technology for AV1}
\newacro{nlp}[NLP]{natural language processing}
\newacro{gpt}[GPT]{generative pre-trained Transformer}
\newacro{llm}[LLM]{large language model}

\usepackage{xspace}

\makeatletter
\DeclareRobustCommand\onedot{\futurelet\@let@token\@onedot}
\def\@onedot{\ifx\@let@token.\else.\null\fi\xspace}

\def\etal{\emph{et al}\onedot}

\title{UAV Immersive Video Streaming: A Comprehensive Survey, Benchmarking, and Open Challenges}

\begin{document}
\author{\IEEEauthorblockN{Mohit K. Sharma, Chen-Feng Liu, \IEEEmembership{Member, IEEE}, Ibrahim Farhat, Nassim Sehad, Wassim Hamidouche,\\and M{\'e}rouane Debbah, \IEEEmembership{Fellow, IEEE}}
\thanks{Mohit Sharma, Chen-Feng Liu, Ibrahim Farhat, and Wassim Hamidouche are with the Technology Innovation Institute, Masdar City, Abu Dhabi 9639, United Arab Emirates (e-mail: \href{mailto:firstname.lastname@tii.ae}{firstname.lastname@tii.ae}).}
\thanks{Nassim Sehad is with Aalto University, 02150 Espoo, Finland (e-mail: \href{mailto:nassim.sehad@aalto.fi}{nassim.sehad@aalto.fi}).}
\thanks{M{\' e}rouane Debbah is with Khalifa University, Abu Dhabi, United Arab Emirates (e-mail: \href{mailto:merouane.debbah@ku.ac.ae}{merouane.debbah@ku.ac.ae}).}}

\maketitle

\begin{abstract}

Over the past decade, the utilization of \acp{uav} has witnessed significant growth, owing to their agility, rapid deployment, and maneuverability. Among various applications, capturing videos using \acp{uav} has emerged as a prominent and promising use case, enabling diverse applications such as remote surveillance and gaming. In particular, the use of \ac{uav}-mounted 360-degree cameras to capture omnidirectional videos has enabled truly immersive viewing experiences with up to \ac{6dof}. However, achieving this immersive experience necessitates encoding omnidirectional videos in high resolution, leading to increased bitrates. Consequently, new challenges arise in terms of latency, throughput, perceived quality, and energy consumption for real-time streaming of such content. This paper presents a comprehensive survey of research efforts in \ac{uav}-based immersive video streaming, benchmarks popular video encoding schemes, and identifies open research challenges. Initially, we review the literature on 360-degree video coding, packaging, and streaming, with a particular focus on standardization efforts to ensure interoperability of immersive video streaming devices and services. Subsequently, we provide a comprehensive review of research efforts focused on optimizing video streaming for time-varying \ac{uav} wireless channels. Additionally, we introduce a high-resolution 360-degree video dataset captured from \acp{uav} under different flying conditions. This dataset facilitates the evaluation of complexity and coding efficiency of software and hardware video encoders based on popular video coding standards and formats, including \acs{avc}/H.264, \acs{hevc}/H.265, \acs{vvc}/H.266, \acs{vp9}, and AV1. Our results demonstrate that \acs{hevc} achieves the best trade-off between coding efficiency and complexity through its hardware implementation, while AV1 format excels in coding efficiency through its software implementation, specifically using the libsvt-av1 encoder. Furthermore, we present a real testbed showcasing 360-degree video streaming over a \ac{uav}, enabling remote control of the drone via a 5G cellular network. Finally, we discuss open challenges and outline future research directions for efficient and low-latency immersive video streaming using \ac{uav}.
\end{abstract}

\begin{IEEEkeywords}
UAV, 360$^\circ$, low latency, video coding. 
\end{IEEEkeywords}
\section{Introduction}
\IEEEPARstart{I}{mmersive} video technology enables users to experience a quasi-realistic virtual environment, fostering engagement and a sense of presence in a digital space. Various visual media modalities, such as volumetric, light field, and \ac{odv}, have emerged as viable options for delivering an immersive viewing experience~\cite{10089176}. Among these, \ac{odv}, commonly known as 360-degree video, has gained widespread popularity due to the availability of acquisition and display devices, as well as standardization efforts ensuring interoperability. Integrating real-time transmission of 360-degree video using a \ac{uav}-mounted camera enhances the immersive viewing experience by adding an extra degree of freedom through \ac{uav} mobility. This advancement holds promise for diverse applications like remote video surveillance, scientific exploration, autonomous manufacturing assistance, agricultural monitoring, and more. However, this acquisition system for 360-degree video presents new challenges in delivering high \ac{qoe}, primarily due to the limited computational and energy resources of \acp{uav} and the rapid fluctuations of wireless channels. Additionally, the need for high-quality video and ultra-low \ac{e2e} latency becomes crucial to ensure real-time control of the \ac{uav}, especially in dynamic mobility conditions, further amplifying these challenges.
%\IEEEPARstart{I}{mmersive} video technology enables users to perceive a virtual environment with a quasi-realistic experience where the user can interact with the surrounding digital environment, augmenting the feeling of engagement and presence in a virtual space. Different visual media modalities, including volumetric, light field, and \ac{odv}, have emerged as serious candidates for offering an immersive viewing experience ~\cite{10089176}. Among these modalities, \ac{odv} (i.e., 360$^\circ$ video) is the most deployed technology thanks to the availability of acquisition and display devices in the market and the standardization efforts that ensure interoperability of 360$^\circ$ devices and services. On the other hand, real-time videos transmission using a 360$^\circ$ camera mounted on a \ac{uav} offers the user an immersive viewing experience with an additional degree of freedom enabled by the mobility of the \ac{uav}, enhancing many promising applications such as remote video surveillance, scientific discoveries, autonomous manufacturing assistance, agriculture monitoring, etc. Nevertheless, this 360$^\circ$ video acquisition system poses new challenges in providing a high \ac{qoe}, stemming mainly from \ac{uav}'s restricted computation and energy resources and rapid wireless channel variations. In addition, high quality and ultra-low \ac{e2e} latency requirements to ensure real-time control of the \ac{uav}, especially in mobility conditions, further exacerbates these challenges.    

Addressing the above challenges will require efforts to enhance the communication for \acp{uav} and develop adaptive and low-complexity schemes for 360$^\circ$ video encoding and streaming. In Table~\ref{tab:table_survey}, we present a summary of recent efforts \cite{8660516, Wu_JSAC_Oct2021, Hayat_IEEECommSurvey_Dec2016, Baltaci_CommSurvey_April2021, Mishra_IEEECommstandardMag_May2020, Yan_IEEE_acess_Jul2019, Zhenyu_IEEEComm_survey_Jan2022, Yongs_IEEEProc_Dec2019, Elmokadem_Sensors_2021, Fotouhi_IEEEComSurvey_Dec2019, Marojevic_VT_Magazine_Feb2020, Abdalla_IEEE_commstandardMag_May2021, Wang_IEEECommSurvey_Feb2020, Yaqoob_IEEEComSurvey_2020, Dongbiao_ICETE_2018, NGUYEN2023103564, Zink_IEEEProc_2019} surveying the state-of-the-art research on communication for \acp{uav} and immersive streaming. The literature in Table~\ref{tab:table_survey} can be broadly classified into two categories: covering the communication aspects of \acp{uav} or the streaming of 360$^\circ$ videos. The authors in  \cite{8660516, Wu_JSAC_Oct2021, Hayat_IEEECommSurvey_Dec2016} presented a comprehensive survey of challenges and fundamental tradeoffs in designing wireless networks involving the \acp{uav}. In particular, Mozaffari \etal~\cite{8660516} described various analytical frameworks and tools to address the design challenges, and Hayat \etal~\cite{Hayat_IEEECommSurvey_Dec2016} surveyed the quality of service, connectivity, safety, and other general networking requirements for unmanned aircraft systems in civilian applications. Similarly, Baltaci \etal~\cite{Baltaci_CommSurvey_April2021} reviewed the connectivity requirements for aerial vehicles, especially for piloting applications, and advocated achieving these stringent connectivity requirements through multi-technology heterogeneous networks. In \cite{Wu_JSAC_Oct2021, Mishra_IEEECommstandardMag_May2020}, the authors evaluated various enabling 6G technologies highlighting their benefits and drawbacks regarding their integration in a 6G wireless network with \acp{uav}. They also discussed the design issues associated with integrating the \acp{uav} into the wireless networks in the presence of these technologies. Further, authors in \cite{Yan_IEEE_acess_Jul2019} surveyed the channel models for air-to-ground and air-to-air channel models for \ac{uav} communication.  

\begin{table}[t!]
    \centering
    \caption{Summary of the State of the Art}
     \resizebox{0.48\textwidth}{!}{%
    \begin{tabular}{cc}
    \toprule
       Category & Summary \\
        \midrule
       \rotatebox[origin=c]{90}{Communication for UAVs}  & \begin{tabular}{lc}
    
       \hspace{55pt} Covered topics  & References  \\
        \midrule
        \vspace{5pt}
        \begin{tabular}{p{6cm}}
         \tabitem Fundamental design tradeoffs\\
         \tabitem Connectivity requirements          
        \end{tabular}
        & ~\cite{8660516, Baltaci_CommSurvey_April2021, Hayat_IEEECommSurvey_Dec2016}   \\
        \midrule
        \vspace{5pt}
        \begin{tabular}{p{6cm}}   
        Enabling technologies for \ac{uav}'s integration into 6G networks:\\
        \tabitem Intelligent reflecting surfaces\\
        \tabitem \ac{mmwave} connectivity\\
        \tabitem Short-packet communication\\
        \tabitem Integrated communication and sensing
        \end{tabular} & 
        \cite{Wu_JSAC_Oct2021, Mishra_IEEECommstandardMag_May2020}   \\
        \midrule
        \vspace{5pt}
         Wireless channel model for \ac{uav}-to-ground channel and vice versa & \cite{Yan_IEEE_acess_Jul2019}  \\
        \midrule
        \vspace{5pt}
        \ac{mmwave}-enabled \ac{uav} wireless networks  &  \cite{Zhenyu_IEEEComm_survey_Jan2022}  \\ 
        \midrule
        \vspace{5pt}
        Joint communication, control, and computation design & \cite{Yongs_IEEEProc_Dec2019, Wang_IEEECommSurvey_Feb2020, Elmokadem_Sensors_2021} \\
        \midrule        \vspace{5pt}
        Standardization, experimentation, and prototyping & \cite{Marojevic_VT_Magazine_Feb2020, Abdalla_IEEE_commstandardMag_May2021, Fotouhi_IEEEComSurvey_Dec2019}  \\
        \midrule
        \vspace{5pt}
        Interference issues in \ac{uav} networks & \cite{Fotouhi_IEEEComSurvey_Dec2019, Yongs_IEEEProc_Dec2019}\\
        \end{tabular}\\
        \midrule 
        \rotatebox[origin=c]{90}{360$^{\circ}$ video streaming} & 
     \begin{tabular}{lc}
     
          Various aspects of 360$^{\circ}$ streaming:\\
          \tabitem 360$^{\circ}$ video delivery architecture\\
          \tabitem Viewport dependent, viewport independent, and tile based solutions
                & \cite{Yaqoob_IEEEComSurvey_2020, Zink_IEEEProc_2019} \vspace{5pt} \\ 
        \midrule
        Issues pertaining to networks for 360$^{\circ}$ streaming & \cite{Yaqoob_IEEEComSurvey_2020, Dongbiao_ICETE_2018} \vspace{5pt} \\
        \midrule
        Compression and coding for 360$^{\circ}$ streaming   & \cite{Dongbiao_ICETE_2018} \vspace{5pt} \\
        \midrule
        Video streaming (\ac{2d}) from aerial platforms & \cite{NGUYEN2023103564} \vspace{5pt} \\
        \end{tabular}\\
        \bottomrule
    \end{tabular}}
    
    \label{tab:table_survey}
\end{table}

To meet high data rate requirements for \acp{uav}, Xiao \etal~\cite{Zhenyu_IEEEComm_survey_Jan2022} reviewed relevant antenna structures and channel models for \ac{mmwave}. Furthermore, the technologies and solutions for \ac{uav}-connected \ac{mmwave} cellular networks and \ac{mmwave}-\ac{uav} ad hoc networks were discussed. The authors in \cite{Yongs_IEEEProc_Dec2019} and \cite{Elmokadem_Sensors_2021} reviewed the methods for communication and trajectory co-design. In addition, Zeng \etal~\cite{Yongs_IEEEProc_Dec2019} surveyed various techniques to deal with the air-to-ground interference issues in cellular communication with \acp{uav}. Fotouhi \etal~\cite{Fotouhi_IEEEComSurvey_Dec2019} investigated the interference issues and potential solutions addressed by standardization bodies for serving aerial users with the existing terrestrial \acp{bs}. In addition, they reviewed the ongoing prototyping, testbed activities, and regulatory efforts to manage the commercial use of \acp{uav}, along with cyber-physical security of \ac{uav}-assisted cellular communication. In \cite{Marojevic_VT_Magazine_Feb2020}, Marojevic \etal presented the architecture of aerial experimentation and research platform for advanced wireless, which facilitates experimental research in controlled yet production-like environments. In \cite{Abdalla_IEEE_commstandardMag_May2021}, Abdalla \etal~surveyed the ongoing \ac{3gpp} standardization activities for enabling networked \acp{uav}, requirements, envisaged architecture, and services provided by \acp{uav}. The authors in \cite{Wang_IEEECommSurvey_Feb2020} studied the \ac{uav} networks from the perspective of cyber-physical systems and considered the joint design of communication, computation, and control to improve the performance of \ac{uav} networks. %They also highlighted the coupling effect among these components by classifying them into three hierarchies, i.e., cell level, system level, and system of systems. In \cite{Elmokadem_Sensors_2021}, the authors reviewed advanced methods capable of producing three-dimensional avoidance maneuvers and safe trajectories. 

On the other hand, the work in  \cite{Yaqoob_IEEEComSurvey_2020, Dongbiao_ICETE_2018, NGUYEN2023103564} surveyed the adaptive streaming techniques for 360$^\circ$ videos. Yaqoob \etal~in \cite{Yaqoob_IEEEComSurvey_2020} reviewed the adaptive 360$^\circ$ video approaches that dynamically adjust the size and quality of the viewport. In addition, they surveyed the standardization efforts for 360$^\circ$ video streaming, highlighting the main research challenges such as viewport prediction, \ac{qoe} assessment, and low latency streaming for both the on-demand and live 360$^\circ$ video streaming. Further, \cite{Dongbiao_ICETE_2018} surveyed the \ac{fov} prediction methods, compression, and coding schemes for reducing the bandwidth required for streaming immersive videos. In addition, they reviewed caching strategies and datasets for immersive video streaming. The work in \cite{NGUYEN2023103564} focused on the issues pertaining to \ac{2d} video streaming from an aerial platform. In particular, they surveyed the works using \ac{ai}-based techniques for enhancing video transmission performance. 

We emphasize that none of the above papers have explicitly surveyed the issues relevant to immersive video streaming from a \ac{uav} platform, which poses unique challenges regarding computational complexity, quality, and \ac{e2e} latency. In this work, we present a survey of solutions related to state-of-the-art schemes for immersive video streaming from a \ac{uav} and benchmark the existing video encoding schemes. In this direction, our contributions are the following:

\begin{itemize}
    \item Review exiting wireless communication techniques for video streaming using \ac{uav}.   
    \item Build a new dataset of immersive 360$^{\circ}$ videos captured from \ac{uav} in different acquisition conditions and scenes.
    \item Assess the coding efficiency and complexity of software and hardware encoders of five video standards and formats for immersive 360$^{\circ}$ video streaming. 
    \item Discuss open challenges related to \ac{odv} streaming over \ac{uav}. 
\end{itemize}

The rest of this paper is organized as follows. Section~\ref{sec:odvstr} presents the main components of the \ac{odv} streaming chain, including acquisition, encoding, packaging, rendering, and optimization. Then, the key performance metrics and wireless optimization techniques for \ac{uav} 360$^\circ$ video streaming are presented in Sections \ref{sec:kpi} and~\ref{sec:uavwrlopt}, respectively. In Section~\ref{sec:360data}, first the proposed \ac{uav} 360$^\circ$ video dataset is presented, and then benchmark and analysis of software and hardware encoders of five video standards are provided in Section~\ref{sec:bench}. Next, the challenges of \ac{odv} streaming over \ac{uav} are discussed in Section~\ref{sec:chal}. Finally, Section~\ref{sec:con} concludes the paper.

\section{Omnidirectional Video Streaming}
\label{sec:odvstr}
\begin{figure}[t]
\centering
\includegraphics[width=\linewidth]{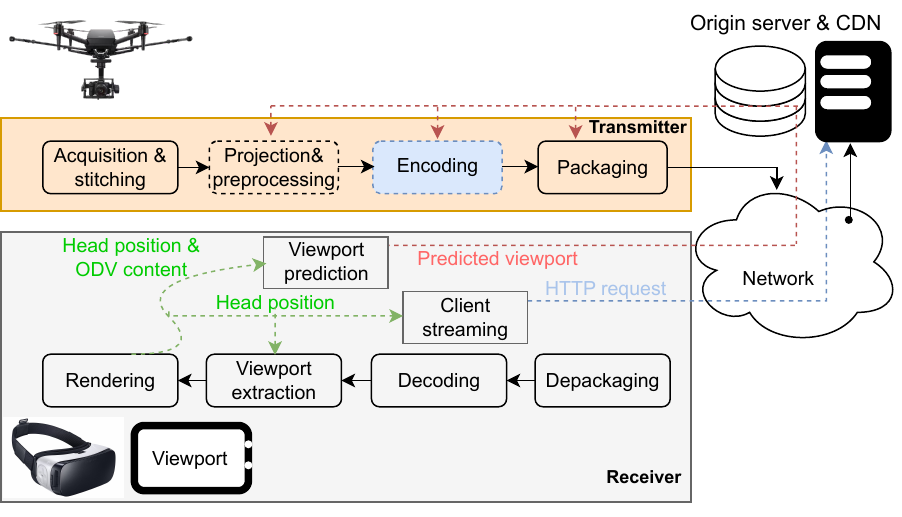}
\caption{ODV \ac{e2e} streaming pipeline.}
\label{fig:vr-stre}
\end{figure}

Figure~\ref{fig:vr-stre} illustrates the \ac{e2e} \ac{odv} streaming pipeline. In this section, we briefly review the technology used at the different stages to deliver \ac{odv} to the end user over the network.
\subsection{Acquisition and Preprocessing}
An omnidirectional visual signal is presented in a spherical space with angular coordinates: the azimuth angle $\phi \in [ \pi, - \pi ]$, and the elevation or polar angle $\theta \in [- \frac{\pi}{2}, \frac{\pi}{2}]$, assuming a unit sphere (radius $r=1$) for acquisition and rendering. The sphere's origin represents the viewing reference that captures the light coming from all directions. %In practice, an omnidirectional visual signal is captured by a multi-view wide-angle acquisition system. The wide-angle capture relies, for instance, on fish-eye lenses. One eye-fish camera enables only a partial sphere capture, while multiple eye-fish camera acquisitions are combined to cover the whole sphere. This operation is performed by {\it stitching} images from different cameras into the sphere~\cite{9633229}. However, the {\it stitching} operation raises two main challenges. The first one is blending and wrapping the non-overlapped captured images. Moreover, inconsistencies in illumination and color that may appear after {\it stitching} need to be corrected. The second challenge appears for video signal, where the camera sensors should be perfectly synchronized. The omnidirectional visual signal in spherical representation is then mapped into a \ac{2d} texture signal in the pre-processing stage before being encoded by conventional \ac{2d} video coding standards. \Ac{erp} is the most commonly used bijective mapping technique, particularly adapted for production and contribution. Yet, more advanced mapping techniques, such as \ac{eac}, \ac{cmp}, and \ac{tsp}, are proposed in the literature~\cite{8902161}. In particular, \ac{cmp} and \ac{tsp} contribute to enhance the coding efficiency by respectively 25\% and 80\% of bitrate savings compared to \ac{erp} and thus are more suitable for distribution~\cite{360meta}. 
In practical applications, an omnidirectional visual signal is captured using a multi-view wide-angle acquisition system, often utilizing fish-eye lenses. While a single eye-fish camera can only capture a partial sphere, combining multiple acquisitions from such cameras allows for complete sphere coverage through the process of {\it stitching} the images together \cite{9633229}. However, the {\it stitching} operation introduces two main challenges. The first challenge involves blending and wrapping non-overlapping captured images, while also addressing inconsistencies in illumination and color that may arise after stitching. The second challenge arises when dealing with video signals, as the camera sensors need to be perfectly synchronized. To facilitate further processing, the omnidirectional visual signal in spherical representation is then mapped to a 2D texture signal during the pre-processing stage, prior to encoding using conventional 2D video coding standards. The most commonly used mapping technique is known as \ac{erp}, which is particularly well-suited for production and contribution purposes. However, more advanced mapping techniques, such as \ac{eac}, \ac{cmp}, and \ac{tsp}, have been proposed in the literature \cite{8902161}. Notably, \ac{cmp} and \ac{tsp} offer enhanced coding efficiency, achieving bitrate savings of 25\% and 80\%, respectively, compared to \ac{erp}, making them more suitable for distribution purposes \cite{360meta}.

\subsection{Encoding}  
After mapping the sphere in a \ac{2d} plane, \ac{odv} content is encoded in practice by conventional \ac{2d} video standards such as \ac{avc}/H.264~\cite{1218189}, \ac{hevc}/H.265~\cite{6316136} \ac{vvc}/H.266~\cite{9503377}, as well as \acs{vp9} and \ac{av1} video formats. In particular, tailored coding tools are integrated into the \ac{hevc}/H.265, \ac{vvc}/H.266 standards to enhance the \ac{odv} coding efficiency and enable advanced streaming features, improving the user's \ac{qoe}. Yet, some non-normative coding techniques were proposed in the literature, encoding the \ac{odv} content in spherical representation to prevent projection distortions, resulting in higher coding efficiency.
%\\\\
%{\bf H.265/\ac{hevc} tools for \ac{odv}}. 
\subsubsection{\ac{hevc}/H.265 Tools for \ac{odv}}
The tile concept in \ac{hevc}/H.265 plays a crucial role in enabling independent and parallel encoding/decoding of rectangular regions within the picture. By breaking the dependency of context prediction in arithmetic encoding and intra prediction, tiles allow for efficient processing and coding of specific regions~\cite{6547985}. Additionally, the tile boundaries also enable the possibility of disabling in-loop filters, further enhancing the flexibility of the encoding process. Moreover, the introduction of the \ac{mcts} technique in \ac{hevc}/H.265, along with \ac{sei} messages, extends the tile concept to the sequence of frames. This technique restricts the \acs{mv} to a selected set of tiles in the reference picture, thereby enabling the download and decoding of only the tiles within the displayed viewport during ODV streaming. This approach significantly improves the user's \ac{qoe} by delivering high-quality content while efficiently utilizing bandwidth.
However, the limitation of restricting \acp{mv} within a set of tiles in the reference picture can have a negative impact on coding efficiency. To overcome this, the literature proposes non-normative solutions that enhance inter-prediction by utilizing the base layer as a reference in the scalable \ac{hevc} extension \cite{10.1145/3210445.3210455}. Alternatively, Bidgoli {\it et al.} \cite{9171593} propose an enhanced intra-prediction technique with fine granularity random access capability, allowing end-users to request specific parts of the stream while ensuring efficient intra-coding.
Furthermore, in the context of spherical bitrate allocation, a new \ac{eeo} strategy is proposed in \cite{8926340}. This strategy derives the Lagrangian multiplier at the block level, which is used in rate-distortion optimization. The proposed solution, evaluated with \ac{erp} and \ac{cmp} projection methods, demonstrates significant bitrate gains when compared to the \ac{hevc} reference software encoder~\cite{8926340}.

%\\\\
%Livre 65     
%{\bf \ac{avc}/H.264 adaptations for \ac{odv}}. There is no tool developed  specifically in the \ac{avc}/H.264 standard for \ac{odv}. Nevertheless, slices can be arranged vertically into a single block column, and their coding constrained and signaled with \ac{sei} messages as for \ac{hevc}. %The picture can also be split prior encoding into sub-pictures, representing a spatial  subset of the original video content. Alternative to slice/tile, the sub-picture sequences are encoded        
%\\\\
%{\bf H.266/\ac{vvc} tools for \ac{odv}.} 
\subsubsection{\ac{vvc}/H.266 Tools for \ac{odv}}
The \ac{vvc}/H.266 standard introduces several advancements for efficient encoding of \ac{odv} content, including the ability to signal the used projection technique and the definition of tailored coding tools~\cite{9503377}. In the case of 360-degree representation and \ac{erp} mapping, objects can span across the left and right picture boundaries continuously. Consequently, in \ac{vvc}/H.266, inter-prediction samples may wrap around from the opposite left or right boundary when \acp{mv} point outside the coded area. Additionally, virtual boundaries are defined to skip in-loop filters across edges. For \ac{cmp} projection, where cube maps may exhibit content discontinuities, virtual boundaries can be signaled to disable in-loop filtering and prevent artifacts arising from non-homogeneous boundaries. Furthermore, \ac{vvc}/H.266 introduces the concept of subpictures, which allows for the extraction of independent rectangular regions within the picture, specifically designed for viewport-dependent \ac{vvc} streaming applications. Subpictures offer two critical improvements over the previous \ac{mcts} concept. Firstly, subpictures enable \acp{mv} to refer to blocks outside the subpicture, and padding at subpicture boundaries is permitted, similar to picture boundaries. This new feature demonstrates higher coding efficiency compared to the tight motion constraints applied in \ac{mcts}. Secondly, the need to rewrite slice headers when extracting a sequence of subpictures to build a new \ac{vvc}/H.266 compliant bitstream is eliminated, streamlining the encoding process~\cite{9503377}.
\subsubsection{Learning-Based Coding for \ac{odv}}
%Machine learning has been widely explored in the literature to optimize and enhance the coding efficiency of \ac{odv}. For example, a \ac{cnn} is trained in~\cite{9000878} to learn the sphere's rotation, which leads to the best coding efficiency. The rotation is performed in pre-processing on the spherical axis before projection, resulting in different rotations of the cube map. The experimental results showed that predicting the rotation enables a substantial coding gain of 8\% to 10\% with a prediction accuracy of 80\%.

%Like conventional video standards, learning-based video codecs can encode the \ac{odv} after its projection on a \ac{2d} plan. The \ac{2d} plan is first transformed into a compact latent space with analysis transform based on an \ac{ann}. The resulting latent representation is then encoded with a lossless entropy encoder to build the bitstream. At the decoder side, the synthesis transform, also based on an \ac{ann}, recovers a reconstructed version of the input \ac{2d} plan from the received bitstream. In addition, the hyper-parameters of the latent space entropy distribution (i.e., mean and variance) are also encoded with an auto-encoder and then leveraged by the encoder and decoder mainly to enhance the performance of the entropy encoder~\cite{minnen2018joint}.    

Machine learning techniques have been extensively investigated in the literature to optimize and improve the coding efficiency of \ac{odv} content. In~\cite{9000878} a \ac{cnn} was trained to learn the rotation of the sphere, resulting in the highest coding efficiency. This rotation is applied as a pre-processing step along the spherical axis before projection, leading to different rotations of the cube map. Experimental results demonstrated that incorporating rotation prediction achieved a significant coding gain of 8\% to 10\% with a prediction accuracy of 80\%.

Similar to conventional video standards, learning-based video codecs can encode \ac{odv} content after its projection onto a 2D plane. Initially, the 2D representation is transformed into a compact latent space using an analysis transform based on an \ac{ann}. The resulting latent representation is then encoded with a lossless entropy encoder to construct the bitstream. At the decoder side, a synthesis transform, also based on an \ac{ann}, reconstructs a version of the input 2D representation from the received bitstream. Moreover, the hyperparameters of the latent space entropy distribution, such as mean and variance, are encoded using an auto-encoder and utilized by the encoder and decoder to enhance the performance of the entropy encoder \cite{minnen2018joint}.
\subsection{Transport Protocols}
Various packaging protocols can be employed for streaming \ac{odv} content, allowing the selection of a suitable protocol based on the specific application and end-user requirements concerning video quality, latency, and advanced functionalities provided by the protocol~\cite{9751563}. In the followin, we outline the key characteristics of two widely utilized streaming protocols: \ac{omaf} and \ac{webrtc}. Specifically, we provide an overview of the primary features of these protocols for ODV streaming. For a more comprehensive understanding, readers are encouraged to consult the comprehensive overview papers on \ac{omaf}~\cite{9380215} and \ac{webrtc}~\cite{7160422}. 
%\\\\
%{\bf \ac{omaf}.} %Introduction 
\subsubsection{\acs{omaf}}
%The ISO/IEC 23090-2 \ac{omaf} is a \ac{mpeg} system standard developed to ensure the interoperability of devices and services targeting storage and streaming of omnidirectional media, including 360$^\circ$ images and video, spatial audio, and associated text. The first version of the standard, finalized in October 2017, provides the essential tools for streaming 360$^\circ$ images and video, enabling \ac{3dof} viewing experience. Additional tools have been integrated into the second version of the standard, released in October 2020, for more advanced features such as enhancing the viewport-dependent streaming, enabling overlays, and multiple viewpoints streaming as the first step towards \ac{6dof} viewing experience. The \ac{omaf} specifications fall within three main modules: content authoring, delivery, and player. Moreover, these specifications are extensions to the \ac{isobmff} and \ac{dash}, ensuring backward compatibility with conventional \ac{2d} media formats.   
The ISO/IEC 23090-2 standard, also known as \ac{omaf}, is a system standard developed by \ac{mpeg} with the objective of ensuring device and service interoperability for storing and streaming omnidirectional media content. This includes various forms of media such as 360$^\circ$ images and videos, spatial audio, and associated text. The initial version of the standard, completed in October 2017, provides fundamental tools for streaming 360$^\circ$ images and videos, enabling a \ac{3dof} viewing experience. In the subsequent release of the standard in October 2020, the second version introduced additional tools to support more advanced features. These features include enhanced viewport-dependent streaming, overlay capabilities, and the ability to stream multiple viewpoints, marking the initial steps towards achieving a \ac{6dof} viewing experience. The specifications of \ac{omaf} are organized into three main modules: content authoring, delivery, and player. Furthermore, these specifications serve as extensions to the \ac{isobmff} and \ac{dash}, ensuring backward compatibility with conventional \ac{2d} media formats.
%Formats 
%\Ac{omaf} supports three omnidirectional visual signal representation types, including {\it projected}, {\it mesh}, and {\it fish-eye}. These formats require different pre-processing for encoding and post-processing for rendering and display. Among the {\it projected} formats, \ac{omaf} supports the two widely used projection algorithms, namely \ac{erp} and \ac{cmp}. Further, a \ac{rwp}  operation enables optional pre-processing operations before encoding, such as resizing, repositioning, or rotation by 90$^\circ$, 180$^\circ$ and 270$^\circ$, and vertical/horizontal mirroring of any rectangular region. The \ac{rwp} operations are conducted for different purposes, such as signaling the exact coverage of partial spherical representation, generating \ac{vs} video, enhancing the coding efficiency, or compensating the over-sampling of pole areas in \ac{erp}. In particular, the \ac{rwp} metadata indicates the applied operations to the player that performs inverse operations to map regions of the decoded picture into the projected picture. 
\ac{omaf} supports three types of omnidirectional visual signal representations: projected, mesh, and fish-eye. Each of these formats requires specific pre-processing for encoding and post-processing for rendering and display. Among the projected formats, \ac{omaf} includes support for two widely used projection algorithms: \ac{erp} and \ac{cmp}. Additionally, \ac{omaf} incorporates a \ac{rwp} operation, which allows for optional pre-processing operations prior to encoding. These operations include resizing, repositioning, rotation by 90$^\circ$, 180$^\circ$, and 270$^\circ$, as well as vertical and horizontal mirroring of specific rectangular regions. \Ac{rwp} serves various purposes, such as signaling the exact coverage of a partial spherical representation, generating \ac{vs} video, enhancing coding efficiency, or compensating for over-sampling in the pole areas of \ac{erp}. The \ac{rwp} metadata indicates the applied operations to the player, which then performs inverse operations to map the regions of the decoded picture back into the projected picture. This ensures proper rendering and display of the content, aligning with the intended transformations specified by the \ac{rwp}.

In addition, the \ac{omaf} standard offers tools for viewport-dependent \ac{odv} streaming, which enables the selection of segments covering the user's viewport at high quality and other segments at lower quality and bitrate. This approach allows for more efficient utilization of network bandwidth, resulting in an improved user experience. Viewport-dependent \ac{odv} streaming can be achieved through two methods: \acl{vs} and tile-based streaming. In the \acl{vs} approach, multiple \acp{vs} are created and signaled, each encoding different viewports at high quality. Users can select the appropriate \ac{vs} stream based on their viewing orientation. The \ac{omaf} \ac{rwqr} metadata can be used to signal the quality of different regions in the sphere. In the tile-based configuration, the \ac{odv} is divided into independent rectangular regions called {\it tiles}. Each tile only depends on the co-located tile in the sequence and can be decoded independently from other tiles. There are two alternatives for encoding video in independent regions. The first method utilizes the \ac{hevc} tile concept, where tiles are grouped into motion-constrained slices known as \acl{mcts}. The second method, applicable to \ac{avc} which does not support tiles, partitions the video into sub-picture sequences, each representing a spatial subset of the original sequence. These sub-picture sequences are then encoded with motion constraints and merged into tiles in a single bitstream. Each tile or sub-picture sequence is stored in its respective track. Additionally, tiles can be encoded in different bitrates and resolutions, allowing users to select the optimal combination of tiles based on their viewing orientation, available bandwidth, and decoding capability.

The \ac{omaf} standard specifies six video media profiles that define the type of video representation and the supported video standard with its associated levels. For example, the "HEVC-based viewport-independent" profile uses the \ac{erp} projected representation and is constrained to \ac{hevc} Main 10 profile level 5.1. This level limits the spatial resolution to 4K (4096 $\times$ 2048). However, the "unconstrained HEVC-based viewport-independent" profile, introduced in the second edition, supports all \ac{hevc} Main 10 profile levels, thus increasing the decoding capacity and display resolution. Furthermore, there are already several open-source implementations available that support the first\footnote{NOKIA: \url{https://github.com/nokiatech/omaf},  Fraunhofer HHI: \url{https://github.com/fraunhoferhhi/omaf.js}, Intel Open Visual Cloud: 
\url{https://github.com/OpenVisualCloud/Immersive-Video-Sample}.} edition of the \ac{omaf} standard. Further, some tools of the \ac{omaf} second edition have been demonstrated in~\cite{10.1145/3339825.3393576, TileMedia1}.

%The \ac{omaf} standard specifies six video media profiles that define the video representation type and the used video standard with its supported levels. For instance, the { \it \ac{hevc}-based viewport-independent} profile uses the \ac{erp} projected representation, and the \ac{hevc} Main 10 profile is constrained to only level 5.1. This latter constrains the spatial resolution up to 4K (i.e., 4096 $\times$ 2048), while the \ac{omaf} { \it unconstrained \ac{hevc}-based viewport-independent} profile, introduced in the second edition,  supports all \ac{hevc} Main 10 profile levels, increasing the decoding capacity and display resolution. Finally, there are already several open-source implementations supporting the first\footnote{NOKIA: \url{https://github.com/nokiatech/omaf},  Fraunhofer HHI: \url{https://github.com/fraunhoferhhi/omaf.js}, Intel Open Visual Cloud: \url{https://github.com/OpenVisualCloud/Immersive-Video-Sample}.} version of the \ac{omaf} standard. Further, some tools of the \ac{omaf} second edition have been demonstrated in \footnote{TileMedia Implementation: \url{ https://www.tiledmedia.com/how-clearvr-drives-and-leverages-standards/}} \cite{10.1145/3339825.3393576}.  
%\\\\
%{\bf \ac{webrtc}.}
\subsubsection{\ac{webrtc}}
The \ac{webrtc} framework is an open-source solution specifically designed to facilitate real-time and low-latency video transmission. Within the \ac{webrtc} transmitter, the "video collector" module takes on the responsibility of video encoding and encapsulating the encoded video frames into \ac{rtp} packets. These packets are subsequently transmitted using the \ac{srtp} protocol. On the receiver side, relevant information regarding the received \ac{rtp} packets is collected, and this information is relayed back to the "video collector" through the transport-wide feedback message of the \ac{rtcp} protocol.
The "bandwidth controller" module, located within the "video collector," utilizes these control messages to compute essential network metrics such as inter-packet delay variation, queuing delay, and packet loss. These metrics play a crucial role in determining the target bitrate, which is then employed by the rate control module of the video encoder. The rate control module dynamically adjusts the encoding parameters, such as the quantization parameter and resolution, based on the target bitrate requirements.
Although the standard \ac{webrtc} implementation does not offer explicit tools for transmitting immersive video, it has gained significant popularity for real-time and ultra-low latency \ac{odv} transmission by treating 360$^\circ$ video representation as a conventional \ac{2d} video \cite{10071541, 10.1145/3394171.3413999}. Additionally, viewport-dependent streaming can be effectively supported by incorporating a combination of high-resolution and low-resolution tiles. This approach optimizes bandwidth utilization while ensuring high quality within the field of view, all while maintaining a low motion-to-photon latency \cite{Intel360}. %\footnote{Intel Advanced 360$^\circ$ Video: \url{https://www.intel.com/content/dam/www/central-libraries/us/en/documents/advanced-360video-implementation-summary-final.pdf}}.
%\ac{webrtc} is an open-source framework designed for real-time and low-latency video transmission. At the \ac{webrtc} transmitter, the {\it video collector} module encodes the video and encapsulates the encoded video frames in \ac{rtp} packets, which are then transmitted through the \ac{srtp}. The receiver collects information on the received \ac{rtp} packets and sends back information to the {\it video collector} in the transport-wide feedback message of the \ac{rtcp}. Based on these control messages, the {\it bandwidth controller} module of the {\it video collector} computes network metrics such as inter-packet delay variation, queuing delay, and packet loss. These metrics are then exploited to calculate the target bitrate used by the rate control module of the video encoder that adapts the encoding parameters (quantization parameter, resolution) according to the target bitrate. Although vanilla WebRTC does not specify tools for transmitting immersive video, it has been widely adopted for real-time and ultra-low latency \ac{odv} transmission by considering the 360$^\circ$ video representation as a conventional \ac{2d} video \cite{10071541, 10.1145/3394171.3413999}. In addition, viewport-dependent streaming can also be supported by mixing high-resolution tiles and low-resolution for efficient bandwidth usage while ensuring high quality in the field of view area along with low motion-to-photon latency.\footnote{Intel Advanced 360$^\circ$ Video: \url{https://www.intel.com/content/dam/www/central-libraries/us/en/documents/advanced-360video-implementation-summary-final.pdf}}
\subsection{Rendering and Display} 
%The limited field of view of the human visual system prevents the end users from directly visualizing the 360$^\circ$ content in the spherical representation. Therefore, only a portion of the sphere (i.e., an image tangent to the sphere called {\it viewport}) is displayed. 
%The visualization requires interaction with the user, creating a sensation of immersion. The interaction consists in whether a head movement (roll, yaw, and pitch),  mouse/keyboard control, or when watching the video from a smartphone, the viewing angle is controlled by moving the device in space, offering a visual experience up to \ac{3dof}. However, the leak of motion parallax feature, which represents the relative position of object changes according to the viewer's position with respect to the object, is one of the main limitations of \ac{odv}. This limitation may lead to discomfort and sickness for the end user. This limitation can be tackled by embedding a 360$^\circ$ camera on a \ac{uav}, providing additional flexibility with mobility to explore the environment and move around the objects of the scene. Therefore, the mobility offered by the \ac{uav} along with the 360$^\circ$ video provides a viewing experience up to \ac{6dof}. Nevertheless, ultra-low \ac{e2e} latency (below 100 ms) and high visual quality are essential to allow interactive control of the \ac{uav}, enabling a more natural viewing experience with accurate control of the \ac{uav}.
The human visual system has a limited field of view, which means that users cannot directly perceive the entire 360$^\circ$ content in its spherical representation. Instead, only a portion of the sphere, known as the "viewport," is displayed, which is an image tangent to the sphere. To enhance immersion, interaction with the user is crucial. This interaction can involve head movements (roll, yaw, and pitch), mouse/keyboard controls, or in the case of viewing on a smartphone, the viewing angle can be controlled by moving the device in space, providing a visual experience of up to \ac{3dof}. However, one of the main limitations of \ac{odv} is the absence of motion parallax, which refers to the relative position of objects changing based on the viewer's position relative to the object. This limitation can lead to discomfort and sickness for users. To address this limitation, a potential solution is to employ a 360$^\circ$ camera mounted on an \ac{uav}. This combination offers enhanced flexibility and mobility, allowing users to explore the environment and move around objects within the scene. By leveraging the mobility provided by the \ac{uav} along with the 360$^\circ$ video, a viewing experience of up to \ac{6dof} can be achieved. However, to enable interactive control of the \ac{uav} and ensure a more natural viewing experience with accurate control, it is essential to have ultra-low \ac{e2e} latency (below 100 ms) and high visual quality. These factors are crucial in order to maintain a seamless and responsive interaction between the user and the \ac{uav}.
\subsection{ODV Streaming Optimization} 
\begin{figure}[t]
    \centering
    %\resizebox{\textwidth}{!}{%
    \begin{forest}
    forked edges,
    for tree={draw,align=center,edge={-latex},fill=white,blur shadow,
        where level=1{
          for descendants={%
           grow'=0,
           folder,
           l sep'+=2.5pt,
           },
        }{}
    }
    [Omnidirectional video streaming, fill=gray!5 
     [Viewport-independent, draw=green!80, fill=green!10
     ]
     [Viewport-dependent
      [Tile-based, draw=red!80, fill=red!10
        [Client-based tile binding]
        [Author-based tile binding]
      ]
      [Projection-based, draw=blue!80, fill=blue!10
      ]
     ]
    ]
    \end{forest}
    %}
 %   \vspace{2mm}
    \caption{\Ac{odv} streaming strategies.}
    \label{fig:cat_dete}
\end{figure}
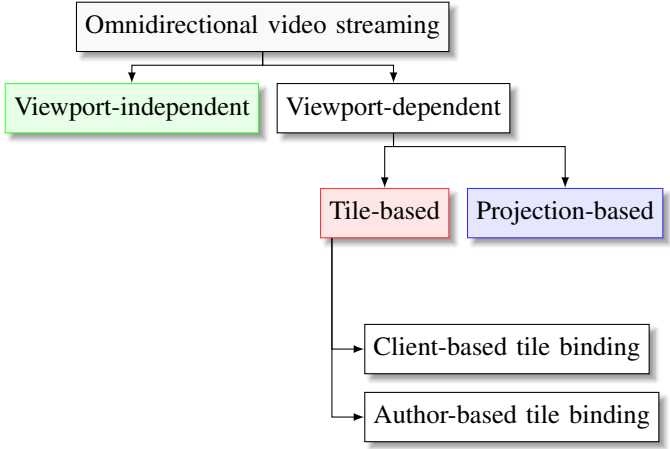

Several strategies have been proposed in the literature to enhance the \ac{qoe} in streaming \ac{odv}~\cite{9830046}. Depending on the specific application, different metrics can be optimized, including perceived video quality, storage cost, bandwidth usage, and various latency measures (e.g., end-to-end, motion-to-photon, or motion-to-high-resolution/quality latency). Low motion-to-photon latency is particularly important to minimize user discomfort when changing the displayed viewport. Additionally, achieving low end-to-end latency is crucial for live \ac{odv} streaming applications, such as \acp{uav}, to enable accurate remote control, especially during high-speed flying conditions. As shown in Figure~\ref{fig:cat_dete}, \ac{odv} streaming strategies can be categorized as either viewport-dependent or viewport-independent, depending on whether the field of view is considered in the optimization process or not. The initial streaming approach, commonly used in early literature, involved transmitting the entire 360-degree content at high quality, allowing users to extract the desired viewport based on their head position with ultra-low motion-to-high-resolution latency. This approach aligns with the viewport-independent profiles defined in the \ac{omaf} standard. However, it is a bandwidth-intensive solution, requiring over 100 Mbps to transmit an 8K resolution video at high quality~\cite{bonnineau2022perceptual}. This is inefficient since the end user only observes a small portion (15\%) of the \ac{odv}. To address this limitation, more advanced techniques have been proposed to provide users with \ac{odv} services at high quality and low motion-to-photon latency. In this context, viewport-dependent strategies have gained wide adoption at the projection (projection-based) and encoding (tile-based) stages. The projection-based approach employs dynamic projection methods, such as pyramidal projection and its refined version, offset cubic projection \cite{10.1145/3083187.3083190}. Offset cubic projection allocates higher pixel density and better quality near the offset direction, which corresponds to the user's viewing direction. Another solution proposed in \cite{10.1145/3394171.3413999} is oriented projection for real-time 360-degree video streaming using the \ac{webrtc} framework. Oriented projection allocates more pixels in the projected frame to areas on the sphere that are close to a target pixel-concentration orientation. This solution is jointly optimized with adaptive resolution and bitrate allocation to accommodate bandwidth variations.

The viewport-dependent \ac{omaf} profiles, as outlined in~\cite{10.1145/3083187.3083190}, enable independent tile decoding by restricting motion vectors to refer only to the adjacent tile area. This allows the end user to request and decode tiles independently. Following the projection stage, the \ac{odv} is encoded into tiles representing different quality representations. The end user can then request the tiles covering the viewport at high quality while the remaining area tiles can be requested at a lower quality. As a result, the tile-based \ac{odv} encoding using viewport-dependent \ac{omaf} profiles significantly improves the user's \ac{qoe} and reduces the required transmission bandwidth. However, storing and encoding \ac{odv} tiles in multiple representations, each with different rate-quality characteristics, necessitates a large storage capacity and incurs high encoder computational complexity. To address these challenges, the \acs{vr} Industry Forum guidelines \cite{VRIndForum} introduced the \ac{hevc}-based \ac{fov} Enhanced Video Profile. This profile employs \ac{hevc} encoding to achieve low-quality coverage of the entire 360-degree video, while high-quality sub-pictures are encoded to cover specific regions of the video. Each bitstream is then encapsulated within a track compliant with the \ac{hevc}-based viewport-dependent \ac{omaf} profile. The player can subsequently request the bitstream covering the viewport in high quality, along with the low-quality bitstream representing the entire 360-degree coverage. In live scenarios, the low-quality stream can be transmitted via multicast, allowing for more efficient bandwidth utilization. This approach ensures efficient bandwidth utilization while maintaining ultra-low motion-to-photon latency. 

The prediction of end user head movements can be leveraged to enhance the \ac{qoe} by assigning higher fetching priority to tiles within the predicted viewport. This streaming approach, known as the "human-centric" approach, focuses on optimizing the user experience, in contrast to the "system-centric" approach that prioritizes overall system performance without considering user behavior. The design can be categorized as single-user or cross-users, with the latter considering the behavior of multiple users in predicting the viewport. These techniques rely on accurate viewport prediction models, which are used to optimize the streaming system. In \cite{10.11452980055.2980056}, the potential of predicting head movements for optimizing 360-degree video streaming over cellular networks was demonstrated, resulting in up to 80\% network bandwidth savings. This approach has been followed by several research papers and commercial products, aiming to optimize network and computational resources while providing users with a highly immersive experience \cite{TileMedia}.

\begin{table}[t!]
    \centering
    \caption{Performance of streaming approaches regarding storage capacity, bandwidth usage, motion-to-photon latency, and encoding time.}
    \label{tab:360streameffi}
    \begin{adjustbox}{max width=0.48\textwidth}
    \begin{tabular}{l|c|c|c|c}
    \toprule
        & Storage & Bandwidth & Latency & Encoding time     \\  \midrule
Viewport-indep.  &   \blackbullets{3} & \blackbullets{1} & \blackbullets{3}  & \blackbullets{3}  \\  \midrule
     Projection-based  & \blackbullets{2} & \blackbullets{2} & \blackbullets{2} & \blackbullets{2}  \\  \midrule
         Tile-based & \blackbullets{1} & \blackbullets{3} & \blackbullets{2} & \blackbullets{1}  \\  \bottomrule
    \end{tabular}
   \end{adjustbox}
   {\begin{flushleft}
   ~ ~ ~ Performance metrics: High $\equiv$ \blackbullets{3}, Average $\equiv$ \blackbullets{2}, Low $\equiv$ \blackbullets{1}
   \end{flushleft}}
\end{table}

Finally, Table~\ref{tab:360streameffi} depicts the performance of discussed 360$^\circ$ video streaming strategies regarding storage cost, bandwidth usage, motion-to-photon latency, and encoding time.

%\begin{figure}[t]
%\centering
%\includegraphics[width=\linewidth]{Figures/wireless_organization.pdf}
%\caption{The organization of Section \hl{Brahim:I think we can remove this}\ref{sec:wrlopt}.}
%\label{fig:wireless}
%\end{figure}

\section{\ac{uav}-Based Real-Time Video Streaming: Performance Metrics}
\label{sec:kpi}
This section focuses on the optimization of wireless networks for real-time video streaming using \acp{uav}, with a particular emphasis on key performance metrics. The discussion encompasses three essential aspects: latency, video quality, and \ac{uav} energy consumption.
\subsection{Latency}
The latency in video transmission from \acp{uav} significantly impact user's \ac{qoe} in 360-degree video streaming, including \ac{e2e} latency. It is captured using several metrics such as motion-to-photon latency, and motion-to-high resolution latency. \\\\
\textbf{End-to-end latency:}
In a point-to-point real-time video transmission, \ac{e2e} latency plays a vital role in ensuring a seamless and immersive experience. It encompasses the total delay from event capture by the sensor to processing, transmission, and actuator response. The \ac{e2e} latency between camera and user's display is often referred to as \ac{g2g} latency. It measures the time difference between when the photons of an event first pass through the camera lens and when the event is displayed to the viewer. Additionally, \ac{g2a} latency represents the time gap between the photon corresponding to an event passing through the camera lens and the availability of the first image corresponding to that event for processing before display. \ac{g2a} latency is crucial in applications utilizing computer vision algorithms for tasks such as control, object detection, segmentation, viewport prediction, and more. Figure~\ref{fig:vr-stre} provides an overview of \ac{g2g} latency and its relationship to \ac{g2a} latency by removing the latency introduced during the display process.

The overall \ac{g2g} latency can be expressed as the sum of delays incurred in camera acquisition, encoding, network transmission, decoding, and display processing. Table~\ref{tab:G2G_latency_brakeup} presents a breakdown of \ac{g2g} latency for a state-of-the-art \ac{webrtc}-based implementation of an \ac{odv} \ac{e2e} streaming pipeline~\cite{Intel_360_Implementation}. This pipeline transmits $8K$ resolution $360^\circ$ videos captured using an Insta 360 camera to a Samsung S10 client. The latency breakdown in Table~\ref{tab:G2G_latency_brakeup} highlights that the acquisition and stitching process, along with the encoder, contribute to approximately 80\% of the total \ac{g2g} latency. Notably, the latency introduced at the transmitter scales proportionally with the video's resolution. At a high level, the total \ac{g2g} latency comprises network latency and latency stemming from the transmitter and client components.

\begin{table}[t!]
    \centering
    \caption{Glass-to-Glass Latency Brakeup\cite{Intel_360_Implementation}}
     \resizebox{0.48\textwidth}{!}{%
    \begin{tabular}{cc}
    \toprule
       Block & Latency (ms) \\
        \midrule
       \rotatebox[origin=c]{90}{\bf {Transmitter}}  & \begin{tabular}{lc}
    
       %\hspace{55pt} Sub-blocks  &   %\\
        %\midrule
        \vspace{5pt}
       Live Streaming \hspace{155pt}
        & 503   \\
        \midrule
        \vspace{5pt}           
        FFMpeg Decoder
         & 
        568   \\
        \midrule
        \vspace{5pt}
         $360^\circ$ stitching & 28.5  \\
        \midrule
        \vspace{5pt}
        HEVC encoder  &  406  \\ 
        \midrule
        \vspace{5pt}
        Video packetizer & 1.9 \\
        \midrule        \vspace{5pt}
        \bf{Total latency at transmitter} & \bf{1508}  \\
              
        \end{tabular}\\
        
        \hline
        \rotatebox[origin=c]{90}{\bf{Client}} & 
     \begin{tabular}{lc}
          \\
          RTP Packet
          
                & 79 \vspace{5pt} \\ 
        \midrule
        Decoder \hspace{175pt} & 34 \vspace{5pt} \\
        \midrule
        Renderer   & 14 \vspace{5pt} \\
        \midrule
        \bf{Total latency at client} & \bf{127} \vspace{5pt} \\
        \end{tabular}\\
        \hline\\
        \vspace{5pt}
        \bf{Total G2G latency} & \hspace{190 pt}\bf{~1745-1856}\\
        \bottomrule
    \end{tabular}}
    
    \label{tab:G2G_latency_brakeup}
    
\end{table}
Based on the preceding discussion, it can be deduced that reducing latency entails reducing the number of processed pixels throughout the \ac{odv} streaming pipeline, which is primarily determined by the framerate and resolution. Additionally, higher framerates and quality necessitate increased transmission rates, resulting in heavier overheads in terms of transmission delay and transmit power. Conversely, the \ac{e2e} delay increases when the encoding bitrate fails to adapt to wireless channel variations. As a result, efforts to minimize latency have a direct impact on video quality. Hence, the design of wireless communications for \ac{uav}-based \ac{odv} streaming predominantly revolves around maximizing video quality while adhering to a latency constraint. Typically, in wireless optimization problems, this latency constraint is imposed as a \emph{delay outage probability} constraint, representing the probability of packet delay exceeding a predefined delay budget. However, it is important to note that the delay outage probability constraint only encompasses queueing and transmission delays, which constitute only a portion of the overall \ac{g2g} delay. \\\\
%Based on the above discussion, it can be inferred that an effort to reduce the latency will require reducing the number of pixels to be processed across the \ac{odv} streaming pipeline, which, in turn, is determined by the framerate and resolution. Furthermore, a higher framerate and quality lead to higher transmission rate requirements and incur heavier overheads in transmission delay and transmit power. On the other hand, the \ac{e2e} delay increases when the encoding bitrate is not adapted to meet the wireless channel variations. Thus, the efforts to reduce latency directly impact video quality. Therefore, wireless communications design for \ac{uav}-based \ac{odv} streaming generally focuses on maximizing the video quality subject to a latency constraint. Typically, in wireless optimization problems, we note that the latency constraint is imposed as a \emph{delay outage probability} constraint, i.e., the probability that packet delay exceeds a predetermined delay budget. The delay outage probability constraint only captures the queueing and transmission delays, which only constitute a part of \ac{g2g} delay. \\\\
{\bf Motion-to-photon latency:} In the context of \ac{odv} streaming, particularly in viewport-dependent scenarios, it is crucial to consider additional latency metrics that can impact the user's quality of service. Two such metrics are \ac{m2p} latency and motion-to-high resolution latency. \ac{m2p} latency measures the delay required to display the new viewport corresponding to the user's updated viewing direction after head movement. It encompasses the time needed to request and render the viewport aligned with the user's viewing direction. The specific streaming approach and the technology of the \ac{hmd} can influence the motion-to-photon latency. Additionally, recent work presented in \cite{Warburton2022} demonstrates the potential of utilizing head motion prediction algorithms at the end user's side to significantly reduce the motion-to-photon latency. These algorithms can effectively anticipate the user's head movements and optimize the rendering process accordingly.

\subsection{\Ac{odv} quality}
The quality of experience for end users is primarily determined by the perceived video quality and latency. In the context of 2D video, widely used full-reference objective quality metrics include \ac{psnr}, \ac{ssim}, and \ac{vmaf}. These metrics provide a comprehensive assessment of the perceived quality by comparing the original and reconstructed videos. However, for 360-degree video content, specialized quality metrics have been proposed to account for the unique geometrical distortions introduced by the spherical representation. Notable examples include Spherical PSNR (S-PSNR) and weighted to spherically uniform PSNR (WS-PSNR), which are full-reference objective quality metrics specifically developed for 360-degree video content~\cite{7993736}.

As mentioned earlier, video quality is influenced by various factors, including encoding bitrate, frame resolution, frame rate, and the characteristics of the air-to-ground wireless channel. Generally, higher quality and lower distortion can be achieved by using a higher bitrate (or resolution) and benefiting from favorable channel conditions. Consequently, selecting a higher video bitrate is preferable for improved video reconstruction quality. However, it is important to note that bitrate selection not only affects video quality but also impacts latency. A higher bitrate necessitates a more stringent throughput requirement, posing challenges for efficient wireless resource allocation. Therefore, when designing the system, a trade-off must be considered between reconstruction quality/distortion and bitrate selection, while optimizing the provision of wireless resources to meet the selected video bitrate.

%The perceived video quality, along with latency, primarily determine the quality of experience of the end user. Peak Signal-to-Noise Ratio (PSNR), Structural Similarity Index (SSIM), and Video Multi-Method Assessment Fusion (VMAF) are three widely popular full-reference objective quality metrics used to evaluate the perceived 2D video quality. For 360-degree video, tailored quality metrics are proposed that consider the geometrical distortions of the \ac{odv} content. Among these metrics, we can cite the Spherical PSNR (S-PSNR) and weighted to spherically uniform PSNR (WS-PSNR), two full reference objective quality metrics proposed for 360-degree video content \cite{7993736}.

%here we can add more values 
%As discussed in the previous section, video quality is affected by several factors, such as encoding bitrate, frame resolution, frame rate, and air-to-ground wireless channel. Generally, better quality/lower distortion is achieved by a higher bitrate (or resolution) and better channel conditions. Therefore, selecting a high video bitrate is preferred for higher-quality video reconstruction. However, as discussed above, the bitrate affects not only the video quality but also the latency. The higher bitrate imposes a stringent throughput requirement. Hence, considering the trade-off between the reconstruction quality/distortion and bitrate selection, the system design needs to optimize the wireless resource provision for the selected video bitrate.  

\subsection{UAV Energy consumption/ Flight Time}
\Acp{uav} are frequently required to maneuver in three-dimensional space to perform various monitoring and video streaming missions. One crucial consideration in \ac{uav}-based \ac{odv} streaming systems is the energy consumption of \acp{uav} due to their limited energy budget. The energy consumed by a \ac{uav} during movement is referred to as "propulsion energy," which is influenced by the \ac{uav}'s velocity and acceleration. Moreover, when the \ac{uav} hovers at a fixed position while streaming the video, it consumes "hovering energy" \cite{GaoZenWanWuZhaSonQiaJin21,YanYanDinChe22}. Furthermore, as discussed below, the air-to-ground channel between the \ac{uav} and the ground user is implicitly affected by the \ac{uav}'s position in the 3D space. For example, the small-scale fading component of the \ac{uav}-ground wireless channel can be modeled as an "angle-dependent Rician fading channel" with the Rician factors directly proportional to the \ac{uav}-ground elevation angle \cite{You_TWC_Jun2018}. This model captures the fact that as the elevation angle increases, the \ac{uav}-ground link tends to experience less scattering, resulting in a larger \ac{los} component. On the other hand, the large-scale fading component, which includes path loss and shadowing, depends not only on the 3D locations of the \ac{uav} and the ground user but also on the geographic distribution of buildings. In urban areas, the signal propagation of a \ac{uav} flying at a lower altitude may be obstructed by buildings, leading to the shadowing effect \cite{YouZha20}. In contrast, when the \ac{uav} transmits at a higher altitude, it only experiences path loss without any shadowing. However, conducting a comprehensive path-loss measurement for a wide geographic area is infeasible. Therefore, a generic probabilistic aerial-to-ground channel model that statistically incorporates both \ac{los} and \ac{nlos} large-scale fading is considered in \cite{HouKanLar14}. In this model, the probability of experiencing \ac{los} path loss increases as the \ac{uav} raises its altitude or moves closer to the ground user horizontally.

In conclusion, the trajectory and position of a \ac{uav} have an impact on its energy consumption and the quality of the transmitted video. Hence, in the deployment and trajectory design of \acp{uav} for video streaming, the distinctive features of the air-to-ground channel, as well as the propulsion and hovering energy consumption, must be carefully taken into account. The following section provides a more detailed exploration of these design challenges and discusses the current state-of-the-art in each aspect.

\section{Wireless Communications Design for \ac{uav}-Based Real-Time Video Streaming} 
\label{sec:uavwrlopt}
This section provides a comprehensive survey of the ongoing research efforts in the design and optimization of wireless communications for real-time video streaming systems using \acp{uav}. Additionally, we also review the relevant standardization activities conducted by \ac{3gpp}.

\subsection{Quality of Experience Maximization}
To meet the demanding requirements of real-time transmission in high-resolution video streaming systems with superior \ac{qoe}, \ac{uav}-based \ac{odv} streaming systems impose stringent criteria on both throughput and latency performance. The latency of such systems is typically characterized by \ac{e2e} latency or photon-to-motion latency. The inherent randomness of wireless channels poses a significant challenge in achieving desired \ac{qoe}, as fluctuating channel conditions result in unpredictable latency, leading to interrupted or choppy video streaming. Maximizing \ac{qoe} is generally approached as a problem of maximizing \ac{psnr} through optimizing wireless resource allocation, including transmit power, rate, or bandwidth, while adhering to various wireless network and \ac{uav}-imposed constraints. In this section, we survey the state-of-the-art advancements in this area. We note that most of the literature in this area has focused only on transmission of \ac{2d} videos from \acp{uav}.

One notable work by Xia {\it et al.} \cite{XiaWanCheCaoJiaZha20} utilized the internal sensor data of the \ac{uav} for adaptive bitrate selection. They leveraged location, velocity, and acceleration information to predict future throughput and proactively select the video bitrate accordingly. The performance evaluation, conducted using a laptop on the ground and the DJI Matrice 100 drone with an attached Android smartphone in an outdoor environment, employed the IEEE 802.11n protocol. The simulations demonstrated that the selected bitrates effectively adapted to future throughput, maintaining relatively stable video bitrates over time, resulting in a seamless video viewing experience despite channel fluctuations. In another study, Muzaffar {\it et al.}~\cite{MuzYanRafBetCav20} focused on a multicast video streaming framework where a \ac{uav} delivers video to ground users. The proposed approach incorporated feedback from the users to dynamically adjust the transmission rate and video bitrate. The performance evaluation, conducted using the AscTec Pelican drone equipped with a Logitech C920 camera and employing the IEEE 802.11a protocol and \ac{avc}/H.264 video format, investigated throughput, packet loss, and delay. The rate-adaptation approach demonstrated improvements in throughput, latency, and packet loss compared to a constant transmission rate and bitrate baseline, resulting in up to 30\% \ac{psnr} gain. These works represent significant advancements in enhancing \ac{qoe} through adaptive bitrate selection and rate control mechanisms, showcasing the potential of optimizing wireless communications in \ac{uav}-based video streaming systems.

In \cite{Cha19}, the authors addressed the transmission rate allocation problem in a \ac{uav} video streaming system, where multiple \acp{uav} transmit their captured videos to different users. The objective was to minimize the overall reconstruction error of all users by optimizing the transmission rates, subject to the total network channel capacity. The performance evaluation, conducted using \ac{psnr} as a measure of video reconstruction quality, showed that the proposed rate allocation approach achieved a 6 dB gain over the equal allocation baseline. Extending the system model in \cite{Cha19}, \cite{HeXieTia19}, considered a multi-\ac{uav} setup, where \acp{uav} competed for transmission rates by incurring a cost to obtain higher rates. Each \ac{uav} aimed to maximize its utility, comprising \ac{psnr} and cost, by selecting a transmission rate within the network capacity budget. The authors designed a rate allocation algorithm using game theory to address the rate competition among \acp{uav}. Compared to the equal bandwidth allocation baseline, the proposed algorithm increased network utility while considering video quality requirements. In contrast to adapting the bitrate to match the channel fluctuations, another stream of work \cite{ZhaMiaZhaYuFuWu20, ZhaCha22} attempts to maximize the PSNR by using an \ac{svc} based video transmission. In \ac{svc}, the video is encoded into a base layer and $N$ enhancement layers. If the $n$th quality is selected for the streamed video, the base layer and all lower enhancement layers, i.e., $1,\cdots,n-1$, have to be delivered along with the $n$th layer \cite{ZhaMiaZhaYuFuWu20}. Note that more enhancement layers give the better quality of the received video, i.e., the higher \ac{psnr}, but require more transmit power at the \ac{uav}. In \cite{ZhaMiaZhaYuFuWu20}, Zhang \etal~considered a system where a \ac{uav} transmits video to a terrestrial \ac{bs} with \ac{svc}. The objective was to maximize the energy efficiency maximization subject to the \emph{delay outage probability} constraint, i.e., the probability that packet delay exceeds a predetermined delay budget. Energy efficiency is defined as the ratio of the \ac{psnr} to the total power. The optimal solution jointly determines the number of enhancement layers and transmit power. In contrast with the baseline, which randomly selects the number of layers and power, the proposed approach improved the energy efficiency by 40\% and decreased the delay outage probability from 0.3 to 0.05. The work \cite{ZhaCha22} studied a video streaming system in which the base and enhancement layers of the \ac{svc} video are sent from a terrestrial \ac{bs} and the \ac{uav} \acp{bs} with the storage and computation capabilities to the ground users. Each layer of the video can be served by either the terrestrial \ac{bs} or a \ac{uav} \ac{bs}, i.e., the user obtains the layers of the video from various \acp{bs}.  The computation capabilities at the \acp{bs} can be used for video processing, e.g., encoding the video's base layer and enhancement layers. In addition, the \acp{uav} without the storage and computation capabilities act as relays to help the transmission from the terrestrial \ac{bs} to the users. Since the number of enhancement layers affects the video quality, the users desire more enhancement layers. By optimizing over the transmit power and allocated bandwidth of the \ac{bs} and \acp{uav}, the numbers of enhancement layers for the users, the video layer assignment (i.e., from which \ac{bs}), and the \ac{2d} deployment of the \acp{uav}, the objective in \cite{ZhaCha22} was to maximize the sum of all users' \ac{qoe} metrics, e.g., normalized \ac{psnr}, subject to the constraint on the transmission and computation delays. The proposed approach achieved 15\% better \ac{qoe}, i.e., received video quality improvement, than a baseline, where the video layers for the user originate from a single \ac{bs} delivering the highest throughput. In contrast with the other baselines in which the video layers for all users originate from the terrestrial \ac{bs}, and the video transmission is helped by the \ac{uav} relays, the proposed approach could achieve 68\% \ac{qoe} further enhancement. Nevertheless, due to its high computational complexity requirements and lack of broad support by consumer devices, the \ac{svc} based approach is not preferable for real-time video transmission.

\subsection{UAV Deployment and Trajectory Design}
In addition to wireless resource allocation, such as transmit power and bandwidth, the maneuverability of \acp{uav} offers an additional dimension for enhancing video streaming performance, including throughput and latency. By optimizing the \ac{uav}'s location or trajectory in \ac{3d} space, both energy consumption and wireless channel conditions can be improved. 

Guo {\it et al.} \cite{GuoCheHuZhe20} focused on the \ac{3d} trajectory design of a \ac{uav} deployed to inspect multiple facilities and transmit real-time video to a control center. The objective was to minimize the total energy consumption associated with propulsion and hovering. The trajectory between successive facilities directly impacted propulsion energy, while hovering energy depended on the inspection time at each facility, determined by video bitrate and transmission latency. Therefore, a trajectory planning algorithm was proposed in \cite{GuoCheHuZhe20} to minimize total energy consumption, assuming a fixed video bitrate. Simulation results demonstrated that the proposed algorithm significantly reduced the \ac{uav}'s energy consumption and flight time. Moreover, resource allocation in terms of time slots, transmit power, and transmission rate was studied in \cite{ZhaHuWanFanNiy20}, where a \ac{uav} delivered videos to multiple ground users. The trajectory design took into account the propulsion energy consumption. Building upon the work in \cite{GuoCheHuZhe20}, Bur {\it et al.} \cite{BurLiuDenChaZah22} extended the research to collaborative inspection of a fire area by multiple \ac{uav}-users, with the inspected videos sent to a \ac{uav}-\ac{bs}. The optimization involved the transmit power of all \acp{uav}, \ac{3d} trajectories of \ac{uav}-users, and dynamic bitrates of the users' inspected videos. The focus was on \ac{qoe} maximization, which accounted for transmission delay violation and the normalized transmission rate based on the selected video bitrate. Additionally, the transmission rate needed to be sufficient to support the selected video bitrate, considering the trajectories and transmit power of the \acp{uav}. The proposed approach enabled the support of 720p and 1080p videos with an average delay of 0.05 ms, whereas a greedy approach relying on immediate \ac{qoe} decisions only supported 140p video with an average delay of 1.2 ms. Overall, these studies highlight the importance of optimizing \ac{uav} trajectories and resource allocation to enhance video streaming performance, achieving energy efficiency, reduced delay, and improved \ac{qoe}.

Furthermore, Khan {\it et al.} \cite{KhaChaGup20} investigated a \ac{uav}-to-\ac{uav} communication network where \acp{uav} collaboratively streamed video to a ground server. Their approach involved utilizing dual paths for transmitting \ac{svc} video with one enhancement layer. The base layer is sent directly from a \ac{uav} to the ground server via a radio frequency link, while the enhancement layer is relayed to the server by neighboring \acp{uav} using free-space-optical links. The objective was to minimize distortion in the received video by jointly optimizing the bitrates of the base and enhancement layers, the routing path, and \acp{uav} deployment. The optimization was subject to a constraint on propulsion energy consumption and the channel capacity's bitrate limitations. The proposed approach achieved an average \ac{psnr} gain of 6 dB compared to a baseline approach that used dual paths with only radio frequency links, without optimizing the routing path and \acp{uav} deployment. In another study, Zhang { \it et al.} \cite{ZhaHuWanFanNiy20} formulated the user's utility as the normalized transmission rate relative to a predetermined bitrate (considering fairness among users). They aimed to maximize the lowest time-averaged utility among all users by jointly designing trajectories and allocating wireless resources. The proposed approach outperformed three baselines: trajectory optimization, wireless resource optimization, and no optimization. It achieved up to a 3-fold increase in transmission rate. In summary, Khan {\it et al.} explored \ac{uav}-to-\ac{uav} communication networks, demonstrating the benefits of jointly optimizing routing paths, \acp{uav} deployment, and bitrate allocation for enhanced video streaming performance. Zhang {\it et al.} focused on maximizing users' utility through joint trajectory design and resource allocation, achieving significant improvements in transmission rates compared to various baselines.

%Further, Khan \etal~\cite{KhaChaGup20} considered a \ac{uav}-to-\ac{uav} communication network in which the \acp{uav} collaboratively stream the video to the ground server. Specifically, the authors utilized dual paths to send the \ac{svc} video with one enhancement layer. The base layer is sent from a \ac{uav} to the ground server via a direct radio frequency link, and the enhancement layer is relayed to the server by the neighbor \acp{uav} through the free-space-optical links. To minimize the distortion of the received video, the bitrates of the base and enhancement layers, the routing path, and \acp{uav} deployment were jointly optimized, subject to propulsion energy consumption constraint and the channel capacity restricted bitrates. As a result, the proposed approach achieved a 6 dB gain of the average \ac{psnr} compared to the baseline, which merely uses the radio frequency links in dual paths without optimizing the routing path and \acp{uav} deployment. By formulating the normalized transmission rate with respect to the predetermined bitrate (i.e., users fairness) as the user's utility, the work \cite{ZhaHuWanFanNiy20} aimed to maximize the lowest time-averaged utility among all users. Compared with three baselines operating with trajectory optimization, wireless resource optimization, and without optimization, the proposed joint trajectory design and resource allocation approach can achieve up to $3\times$ higher transmission rate. 
%\textcolor{blue}{To Do: Perhaps, it'll be good to present at least one UAV channel model which illustrates this LoS and NLoS aspect.}

\subsection{Control Command and Video Data Coexistence}
In the context of \ac{uav} teleportation, the operator at a remote location guides the \ac{uav} to perform critical missions using the live video feed. This involves simultaneous uplink streaming of real-time video and downlink delivery of control commands. Achieving high-quality video streaming necessitates high throughput and low latency, while delivering control commands requires ultra-reliable and low-latency communication. Hence, in wireless systems designed for \ac{uav} teleportation, ensuring reliable and low-latency control command delivery, as well as optimal throughput and latency performance for video streaming, becomes crucial. In the following sections, we examine recent testbed setups that addressed the co-existence of control command and video data, focusing on reliability and latency performance.

%In \ac{uav} teleportation, a remote user navigates the \ac{uav} to execute critical missions based on the live video from the \ac{uav}. The real-time video streaming and control command delivery proceed simultaneously on uplink and downlink, respectively. While streaming high-quality live videos requires high throughput and low latency, delivering the control command from the user demands ultra-reliable and low-latency communication. Therefore, in the wireless systems for \ac{uav} teleportation, reliability and latency performance in the control command delivery and throughput and latency performance of the video streaming is of paramount importance. In the following, we review recent testbed setups which analyze the control command and video data co-existence issue. 

Stornig et al.~\cite{StoFakHelPopBet21} employed the ns-3 network simulator to study \ac{e2e} delays and video quality metrics (\ac{psnr} and \ac{ssim}) in video streaming over 4G networks. They modeled the \ac{uav}'s \ac{3d} trajectory using a Gauss-Markov mobility model, and the video traffic was simulated using the MPEG-4 format with the Evalvid application. The impact of \ac{uav} mobility on latency performance was thoroughly examined. Simulation results indicated that approximately two-thirds of frames were received with good or excellent quality, while 27\% of frames in regular mobility and 30\% of frames in high mobility exhibited inferior quality. Moreover, the average \ac{psnr} and \ac{ssim} values for the received video were 33 dB and 0.945, respectively, indicating good quality.

In the testbed presented in \cite{ZhoHuJurMehDen21}, a DJI Matrice 100 drone equipped with the Quectel EC25 \ac{lte} module and a Raspberry Pi camera were utilized. A computer with a USRP B210 radio frequency unit served as the \ac{bs}, connected to the remote controller via a wireline connection. The experiments were conducted indoors using the \ac{avc}/H.264 video standard. Various metrics were evaluated, including transmission delay, packet loss probability of control commands, and video data throughput. The results demonstrated that when the control command was updated less than 40 times per second, the command delivery experienced a 20 ms transmission delay without any packet loss. Furthermore, the average delay and throughput for 480p and 720p video resolutions ranged from 1.5 s to 5.5 s and from 2 Mbps to 9 Mbps, respectively. In \cite{JinMaLiuLuWuHuaQin21}, the authors evaluated the performance of a testbed equipped with the Huawei MH5000 5G module, operating in an outdoor environment. The transmission rates for streaming 1080p video in \ac{hevc}/H.265 format over 4G and 5G networks were measured at 16 Mbps and 97 Mbps, respectively. The \ac{g2g} delays were evaluated as 1.2 s and 3 s for the respective networks. Additionally, the \ac{e2e} delay of control command delivery was measured to be 30 ms in the 5G network. In a different study, \cite{9951155}, an immersive \ac{uav} control testbed was implemented using the Oculus Quest 2 \ac{hmd} to control \ac{uav} movement and \ac{fov} over 4G, 5G, and WiFi networks. The Insta360 One X camera captured 360$^\circ$ video, and streaming rates of 2 Mbps to 8 Mbps were considered, investigating various delays: \ac{g2g} delay, glass-to-reaction-to-execution delay, and sensor reaction delay. The \ac{g2g} delay ranged from 0.595 s to 0.985 s, the glass-to-reaction-to-execution delay ranged from 0.89 s to 1.38 s, and the sensor reaction delay ranged from 0.67 s to 1.12 s as the streaming rate varied 2 Mbps to 8 Mbps. The control command transmission delay was measured at 138 ms, 103 ms, and 88 ms for 4G, 5G, and WiFi networks, respectively. Additionally, the \ac{psnr} quality of the received video for 720p and 4K resolutions ranged from 30 to 47 dB.

\subsection{Wireless Network Architecture}
Based on the aforementioned quality factors, the design of wireless systems for \ac{uav}-based video streaming can vary depending on the specific wireless network architectures employed. Each network architecture comes with its own restrictions, advantages, overheads, and hardware requirements, leading to diverse performance outcomes. However, poor performance can significantly hinder the feasibility of real-time \ac{uav} video streaming. Therefore, it is crucial to incorporate practical and distinct features that evaluate video streaming quality and \ac{uav} communication performance in the design of wireless systems for \ac{uav}-based video streaming. The evaluation outcomes can also serve as guidance for selecting the appropriate network, depending on the application requirements of \ac{uav}-based video streaming.

In previous works, such as \cite{QazSidWaq15,NavQazKhaMus19}, the network simulator ns-3 was utilized to investigate the performance of \ac{uav} video streaming in 4G networks. The Evalvid application was employed to simulate the video transmission from the \ac{uav} to the \ac{bs} using MP4 format videos. The study in \cite{QazSidWaq15} primarily focused on throughput investigation in both outdoor and indoor environments. In the outdoor scenario, the average throughput achieved by a static macrocell \ac{uav} was found to be 60 kbps, while the throughput decreased to 20 kbps as the \acp{uav} moved at speeds ranging from 1 to 5 m/s. In the indoor environment, the improvement in throughput was more significant for multi-story buildings with an increased number of deployed femtocell \acp{bs}.

Naveed \etal~\cite{NavQazKhaMus19} explored the relationship between the reference signal received power (RSRP) and throughput. Their findings revealed that as the RSRP varied from -110 dBm to -75 dBm, the \ac{uav} achieved video streaming throughputs ranging between 2 kbps and 80 kbps. Additionally, the authors evaluated the received video quality using \ac{psnr} and \ac{ssim} scores under various wireless channel conditions. The \ac{psnr} scores were observed to be 49.41 dB, 35.42 dB, and 24.31 dB in the best, good, and poor channel conditions, respectively. Similarly, the \ac{ssim} scores were found to be 0.99, 0.63, and 0.35 in the respective channel conditions. Furthermore, the impacts of different channel conditions on video quality were visually highlighted through sample videos.

In another study by Sinha \etal~\cite{SinCha21}, the network simulator ns-2.29 was employed to evaluate the throughput, packet loss, packet retransmission, and \ac{e2e} delay performance of video streaming between \acp{uav} and from the \ac{uav} to the ground control station in different network configurations, including wireless local area network (WLAN), WLAN router, WiFi hotspot, and WiFi Direct. Results indicated that WiFi Direct achieved the best performance for all metrics, followed by the WiFi hotspot, while the WLAN network exhibited the poorest performance in all considered metrics.

The performance evaluation of multi-path video streaming in 4G networks was conducted in the works by Liu \& Jia \cite{LiuJia22} and Nihei {\it et al.} \cite{Nihei22}. In the testbed presented in \cite{LiuJia22}, video data was transmitted from dual devices inside the \ac{uav} to a smartphone. The dual-stream approach employed in this study demonstrated the capability to reduce the \ac{e2e} delay to approximately 50 ms, contrasting the single-stream approach. In an independent study, Nihei {\it et al.} \cite{Nihei22} tested the multi-path video streaming method in 4G networks by distributing the video data over two 4G mobile network operations in Indonesia. The objective of data splitting was to minimize the average \ac{e2e} delay. The experimental setup involved the use of a DJI Spreading Wings S800 drone equipped with a Raspberry Pi camera. Outdoor experiments were conducted using the \ac{avc}/H.264 format, demonstrating the ability to adapt the video data to network throughput. Visual illustrations provided in the study showcased the quality improvement achieved with the multi-path method, highlighting its potential suitability for forest fire surveillance. The performance of 60 GHz \ac{mmwave} for video transmission was evaluated by Yu {\it et al.}~\cite{YuTakKaiSak21}. In their experiment conducted in an outdoor environment, a 4K uncompressed video was transmitted from the \ac{uav} to a nearby server to offload further computations. The testbed achieved a throughput of 1.65 Gbps, and the results indicated that offloading computations to the server enabled the \ac{uav} to save 271.8 watts in computations at the expense of 4.1 watts for \ac{mmwave} communication. In contrast to the aforementioned studies, Hu et {\it al.} \cite{HuDenAgh22} conducted a numerical analysis of a \ac{uav}-based \ac{odv} streaming system, where ground users requested specific video tiles within their \ac{fov} from the \acp{uav}. The \ac{uav} then transmitted the requested tiles to the users via associated \acp{ap}, which acted as decode-and-forward relays. These \acp{ap} collaboratively broadcasted the video data to the corresponding users. The objective of their approach was to maximize the \ac{psnr} by scheduling time slots to the \acp{uav} and associating them with the \acp{ap}. The proposed approach yielded an enhancement in \ac{psnr} compared to baselines where each \ac{ap} worked independently or all \acp{ap} worked together.

It is worth noting that while the majority of the studies discussed in this section focused on non-real-time video streaming, they offer valuable insights into the design of \ac{uav}-based real-time \ac{odv} streaming. For example, the work by Yu {\it et al.} \cite{YuTakKaiSak21} emphasizes the importance of joint communications, computation, and control design for \ac{uav}-based real-time video streaming. Similarly, the results presented in \cite{LiuJia22,Nihei22} demonstrate the effectiveness of multi-path streaming in significantly reducing \ac{e2e} delays. Furthermore, the aforementioned studies shed light on the impact of various network settings on video streaming performance.

\subsection{Overview of Relevant 3GPP Standardization Activities}
The standardization activities related to \ac{uav}-based immersive video streaming within the \ac{3gpp} can be divided into two main categories. The first category involves the integration of \acp{uav} with cellular networks, while the second category focuses on 5G support for media streaming applications, such as augmented reality, virtual reality, and real-time communication. In the following sections, we provide a comprehensive overview of the recent advancements and state-of-the-art developments in these two areas.

\subsubsection{Communication for UAVs}
%To understand the potential of \ac{lte} networks to provide cellular support for \acp{uav}, \ac{3gpp} started the Release 15 study in March 2017 \cite{Lin18}. The outcomes of this study are reported in TR 36.777 \cite{3GPP_TR_36777}. The study identified that due to a higher chance of experiencing line-of-sight signal propagation, severe interference might occur in uplink and downlink \ac{uav} communications. Hence, study-item and work-item phrases propose various interference detection and mitigation solutions. In addition, mobility information management and aerial user identification solutions are proposed. Furthermore, in Release 16, the possibility for remote identification of \acp{uav} is studied \cite{3GPP_TS_22125_Rel16}. To further aid the operations of \acp{uav}, in Release 17, 3GPP explored the possibility of 5G support for \acp{uav}, providing functionalities related to \ac{uav} authentication, authorization and tracking \cite{3GPP_ATIS_Rel17_UAV}. Further, it also allows for command and control authorization.  

To evaluate the potential of \ac{lte} networks in supporting \acp{uav} through cellular connectivity, the \ac{3gpp} initiated the Release 15 study in March 2017 \cite{Lin18}. The findings of this study are documented in TR 36.777 \cite{3GPP_TR_36777}. The study revealed that the line-of-sight signal propagation in \ac{uav} communications increases the likelihood of severe interference in both uplink and downlink scenarios. Consequently, various interference detection and mitigation solutions were proposed as study items and work items. Additionally, solutions related to mobility information management and aerial user identification were put forth. In Release 16, the focus shifted towards investigating the feasibility of remotely identifying \acp{uav} \cite{3GPP_TS_22125_Rel16}. In Release 17, \ac{3gpp} further addressed the operational 5G support of \acp{uav} by providing functionalities for \ac{uav} authentication, authorization, and tracking~\cite{3GPP_ATIS_Rel17_UAV}. Moreover, it allows for command and control authorization.
\subsubsection{Support for Media Streaming over 5G}
The support for virtual reality (VR) over wireless networks was initiated in \ac{3gpp} Release 15 with the publication of TR~26.918 \cite{3GPP_TR_26918_Rel15}. This report aimed to identify the potential gaps and use cases for facilitating \ac{vr} services over wireless networks. In addition, Release 17 TS~26.118 introduced operation points, such as resolution and color mappings, and defined media profiles for the distribution of \ac{vr} content. To address the challenges associated with real-time immersive media streaming, Release 18 of \ac{3gpp} is currently investigating several relevant issues. For a comprehensive overview of the activities under Release 18, please refer to Table~\ref{tab:Rel18_activities}.

 \begin{table}[t]
\centering
 \caption{Summary of 3GPP Release~18 activites for supporting media streaming on 5G networks}
\begin{tabular}{>{\centering}m{2.8cm}>{\centering\arraybackslash}m{4cm}}
\hline
{\bf 3GPP Document} & {\bf Focus} 
\\\hline
TS 26.501\cite{3GPP_TS_26501_Rel18} & 5G Media Streaming (5GMS); General description and architecture 
\\\hline
TS 26.506\cite{3GPP_TS_26506_Rel18} & 5G real-time media communication architecture 
\\\hline
TS 26.522\cite{3GPP_TS_26522_Rel18} & 5G real-time media transport protocol configuration
\\\hline
TS~26.803\cite{3GPP_TS_26803_Rel18} & Study on 5G media streaming extensions for edge processing
\\\hline
TR~26.927\cite{3GPP_TR_26927_Rel18} &   Artificial intelligence and machine learning in 5G media services  
\\\hline
 \end{tabular}
 \label{tab:Rel18_activities}
\end{table}
%\cite{Abdalla_IEEE_commstandardMag_May2021,Fotouhi_IEEEComSurvey_Dec2019,MurLinMaaSedZouHapYas21,BorSalSerYan19}.

%Impact of multimedia protocols on the \ac{e2e} latency not for UAV \cite{CaoSuFinPauAmmHui21}
%For information on multipath via ATSSS.

%UE aggregation and multipath via ATSSS (Access Traffic Steering, Switching
%and Splitting) is part of the 3GPP Rel. 16 specifications. ATSSS introduces the
%notion of a Multi Access PDU session where data traffic can be served over
%one or more concurrent accesses [8]. The extent to which these features are
%applicable for more efficient cellular bonding is to be studied.

%3GPP TS 24.193: “5G System; Access Traffic Steering, Switching and Splitting (ATSSS); Stage 3 (Release 17)”

%3GPP R4-092042 standard

\section{Dataset Description}
\label{sec:360data}
%As discussed in previous sections, visual attention and saliency information can be leveraged to learn about humans' visual scene analysis patterns. This, in turn, can be used for developing efficient encoding and streaming schemes. Visual attention and saliency information can be obtained by analyzing the data about viewers' \ac{hm} and \ac{em} while watching the video. This section presents a survey of \ac{em} and \ac{hm} datasets for \ac{odv} videos captured using \acp{uav} and then introduces our new dataset. 
As mentioned earlier, the utilization of visual attention and saliency information can provide valuable insights into human visual scene analysis patterns. This knowledge can be harnessed to develop effective encoding and streaming methods. Visual attention and saliency information can be derived by analyzing viewers' \ac{hm} and \ac{em} during video playback. In this section, we present a comprehensive survey of existing \ac{em} and \ac{hm} datasets for \ac{odv} captured by \acp{uav}. Additionally, we introduce a new dataset that we have curated for this study.
\begin{table*}[t]
    \centering
    \caption{Summary of Existing Datasets}
    \begin{tabular}{lcccc}
    \toprule
        Dataset & Resolution & Frame rate & Dimension & Description \\
         \midrule        
        EyeTrackUAV2\cite{Perrin_MDPI_2020} & $1280 \times 720$ and $720 \times 420$ & $30$ fps & $2$D & Eye tracking data \\
          \midrule
        AVS1K\cite{Fu_TIP_2020} & $1280\times 720$ & $30$ fps & $2$D & Eye tracking data \\
          \midrule
        WinesLab \cite{Colonnese_EUVIP_2018} & $1080 \times 1920$ & $30$ fps & 360$^\circ$ & Videos recorded using both handheld and \ac{uav} mounted camera \\
         \midrule
        360 Track\cite{Mi_applied_sciences_2019} &  $3840 \times 2160$  & $30$ fps & 360$^\circ$ & Includes the ground truth for tracking \\ 
         \midrule
        Proposed & $3840 \times 2160$ & $30-50$ fps & 360$^\circ$ & Table~\ref{tab:table_our_dataset} \\
        \bottomrule
    \end{tabular}
    
    \label{tab:table_dataset}
\end{table*}

In the existing literature, several works have introduced datasets focused on \ac{odv}, encompassing \ac{em} and \ac{hm} information of viewers~\cite{David_MMSys_2018}. For better understanding the user behavior while watching \acp{odv}, these datasets categorize the \acp{odv}, based on the number of moving objects and camera motion, and include users' feedback about viewing experience \cite{Nasrabadi_MultiSys_conf_2019}. On the other hand,  \cite{Wu_MMSys_2017} classified the videos based on their genre, such as documentary, movie, etc. The majority of these datasets consists of videos with \ac{3dof} which makes them less suitable for learning the user viewing pattern for a \ac{uav} based \ac{odv} streaming protocol. Indeed, inferences obtained using \ac{odv} with \ac{3dof} may not be applicable for video transmission platforms with \ac{6dof}, such as \ac{uav} based \ac{odv} transmission. This raises the need to develop novel datasets of \acp{odv} captured using \acp{uav}. In the following, we briefly survey the existing datasets based on the videos captured from \acp{uav}. 

While many datasets in the literature include images and 2D videos captured by \acp{uav} for applications such as remote sensing and navigation, only a limited number of publicly available datasets capture \ac{em} and \ac{hm} information for \ac{uav}-recorded videos, with only one dataset currently accessible \cite{Colonnese_EUVIP_2018}. Similarly, there is only one dataset available for \ac{uav}-based 360$^\circ$ videos. We summarize these datasets in Table~\ref{tab:table_dataset}. The EyeTrackUAV2 dataset \cite{Perrin_MDPI_2020} collects binocular gaze information from 30 viewers watching 43 2D videos under both free viewing and task conditions. The AVS1K dataset comprises ground truth salient object regions for 1000 videos observed by 24 viewers in free viewing conditions. The WinesLab dataset contains 11 360$^\circ$ videos, seven of which were recorded by a pedestrian using a handheld camera, while the remaining four were captured using a drone-mounted camera in various surroundings and lighting conditions. The 360Track dataset consists of 9 360$^\circ$ videos with manually marked ground truth positions of salient objects.  Additionally, we describe in the following our dataset of \emph{aerial} 360$^\circ$ videos presented in Table~\ref{tab:table_our_dataset}.

The dataset presented in Table~\ref{tab:table_our_dataset} comprises a total of ten 360-degree videos. The resolution for all videos, except for ``FreeStyleParaGliding," is 3840 $\times$ 1920, while ``FreeStyleParaGliding" has a resolution of $5120\times 2560$. Each video sequence in the dataset has a length of 40 seconds. The majority of videos, except for ``DubaiVertical" and ``AbuDhabiCity," have a frame rate of 30 frames per second (fps), whereas ``DubaiVertical" and ``AbuDhabiCity" consist of 50 fps. The dataset consists of five outdoor videos, one sports video, and one video recorded in night time conditions. The ``NorthPoleTrip" video captures motion in the azimuth plane, while the ``DubaiVertical" video captures motion in the elevation. Lastly, all video sequences are encoded and transmitted using the \ac{erp} representation.

\begin{table}[t]
    \centering
    \caption{Summary of Our Dataset}
    \resizebox{0.48\textwidth}{!}{%
    \begin{tabular}{lccccc}
    \toprule
        \textbf{Sequence Name} & \textbf{Spatial Resolution} & \textbf{\#Frames} & \textbf{Frame rate (fps)}  & \textbf{Scene Feature} \\
         \midrule      
        PetraJordan & 3840 $\times$ 1920 & 1200 & 30   & Outdoor \\
         \midrule
        CapeTownCityPenorama & 3840 $\times$ 1920& 1200 & 30  & Outdoor \\
         \midrule
        CapeTownCityBeach & 3840 $\times$ 1920 & 1200 & 30   & Outdoor \\
        \midrule
        CapeTownCityGarden & 3840 $\times$ 1920 & 1200 & 30  & Outdoor \\
         \midrule
        CapeTownCitySquare & 3840 $\times$ 1920 & 1200 & 30  & Outdoor \\
         \midrule
        FreeStyleParaGliding & 5120 $\times$ 1920 & 1200 & 30   & Sports \\
         \midrule
        StPetersBergMuseum & 3840 $\times$ 1920 & 1200 & 30  & Night \\
         \midrule
        NorthPoleTrip & 3840 $\times$ 1920 & 1200 & 30  & Motion \\
        \midrule
         DubaiVertical & 3840 $\times$ 1920 & 2000 & 50   & Vertical Motion \\
        \midrule           
        AbuDhabiCity & 3840 $\times$ 1920 & 2000 & 50  & City Panorama \\
         \bottomrule
    \end{tabular}}
    \label{tab:table_our_dataset}
\end{table}

%\begin{itemize}
%\item survey existing datasets on ODV and \ac{2d}, specifically for drones
    %\item Propose a new dataset, using 360 degree Youtube videos. Variety in velocity, content, scenarios etc.
   % \item Check on youtube for drone based 360 videos
%\end{itemize}

\section{Benchmark and Analysis}
\label{sec:bench} 
In this section, we first perform a comprehensive performance benchmarking of five video coding standards and formats (i.e., \ac{avc}/H.264, \ac{hevc}/H.265, \ac{vvc}/H.266, \ac{av1}, and \acs{vp9}) through their software implementations: libx264, libx265, \ac{vvenc}, libvpx-vp9, and libsvtav1, respectively. We also considered two NVIDIA hardware encoder designs hevc\_nvenc and avc\_nvenc, for the \ac{avc}/H.264 and \ac{hevc}/H.265 standards, respectively. All encoders are configured in their fastest preset, targeting live 360$^\circ$ video streaming applications. Table \ref{tab:codec} gives the used hardware and software encoder libraries for the five standards and formats. The encoding process was deployed on a DELL precision 7820 tower workstation. This later is equipped with an Intel Xeon CPU with 8 cores running at a maximum frequency of 3.9 Ghz and a NVIDIA RTX A5000 GPU. % (see Table \ref{tab:hw-caract} for details). 
Furthermore, we present a real test-bed for real-time drone \ac{odv} streaming using a hardware \ac{avc}/H.264 encoder and \ac{webrtc} streaming protocol, enabling remote \ac{uav} control and navigation with \ac{6dof} viewing experience. 
\begin{table}[t]
    \centering
    \caption{Parameters of used encoding workstation}
   % \resizebox{0.48\textwidth}{!}
  {%
  \begin{tabular}{lc}
    \toprule
        \textbf{CPU} &   Intel Xeon Silver \\
         \midrule      
    %     Architecture & x86   \\
    %     \midrule
         \#cores &  8 \\
         \midrule
         Max Freq (GHz) &  3.9 \\
         \midrule
      %    HDD &  2 TB@7200 RPM \\
       %  \midrule
           RAM &  32 GB \\
         \midrule
         SSD &  256GB  \\
         \midrule
         \textbf{GPU} &  NVIDIA Ampere RTX A5000 \\
      %   \midrule
     %    Architecture &  Ampere \\
         \midrule
         \textbf{Operating System} &  Ubuntu 20.04 \\
         \bottomrule
    \end{tabular}}
    \label{tab:hw-caract}
\end{table}
\subsection{Coding and complexity performance}
This section evaluates the coding performance and speedup of the considered software and hardware encoders on video contents captured by a \ac{uav}. The quality of decoded 360$^\circ$ videos is assessed using three objective quality metrics: Spherical \ac{psnr} (S-PSNR), \ac{ssim}, and \ac{vmaf}.  The videos are encoded at four practical \ac{uav} target bitrates of  1.5  Mbps, 3 Mbps, 4.5 Mbps, and  5.8 Mbps, enabling the computation of the BD-rate performance. The BD-rate gives the average bitrate saving or loss compared to the anchor encoder over the four considered bitrates.      

\begin{figure*}[h]
    \centering
    \subfigure[S-PSNR]{\includegraphics[width=0.32\textwidth]{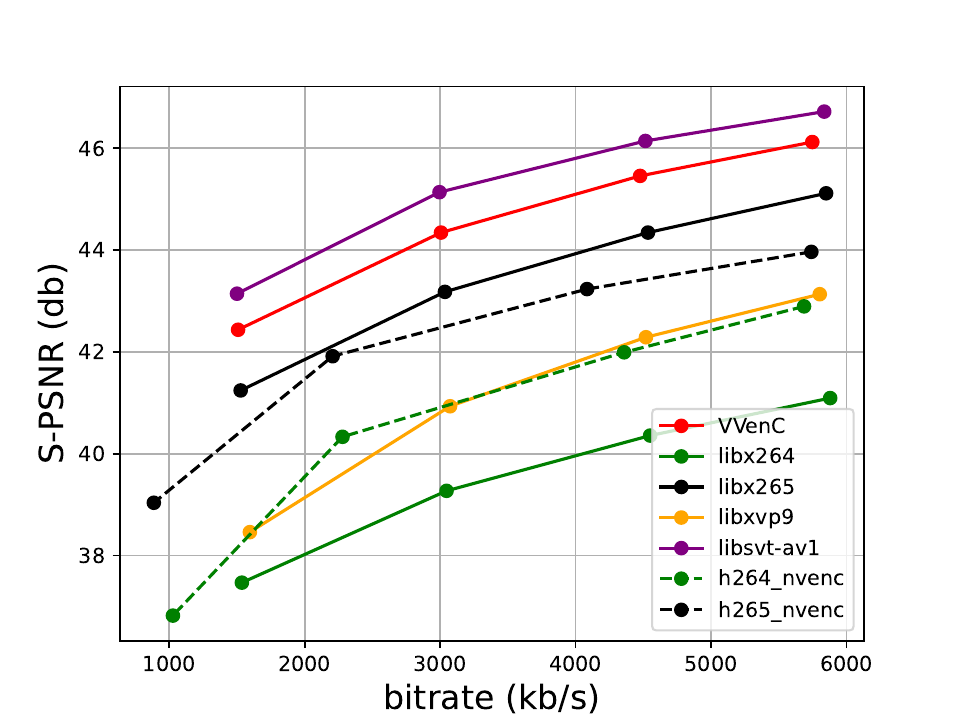}\label{fig-codec_res1a}}
    \subfigure[SSIM]{\includegraphics[width=0.32\textwidth]{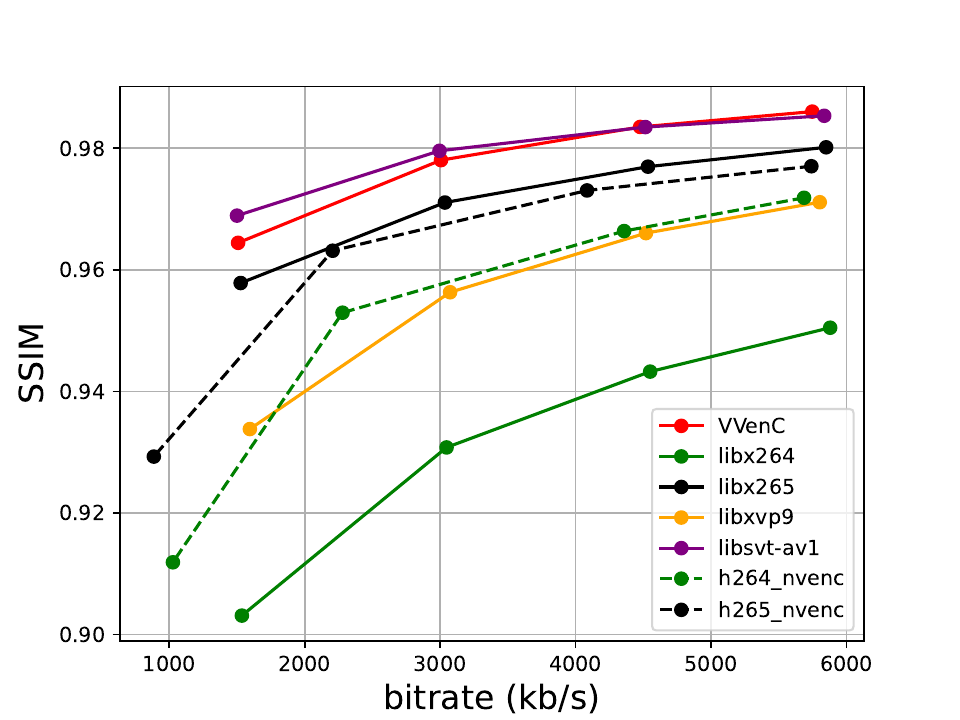}\label{fig-codec_res1b}}
    \subfigure[VMAF]{\includegraphics[width=0.32\textwidth]{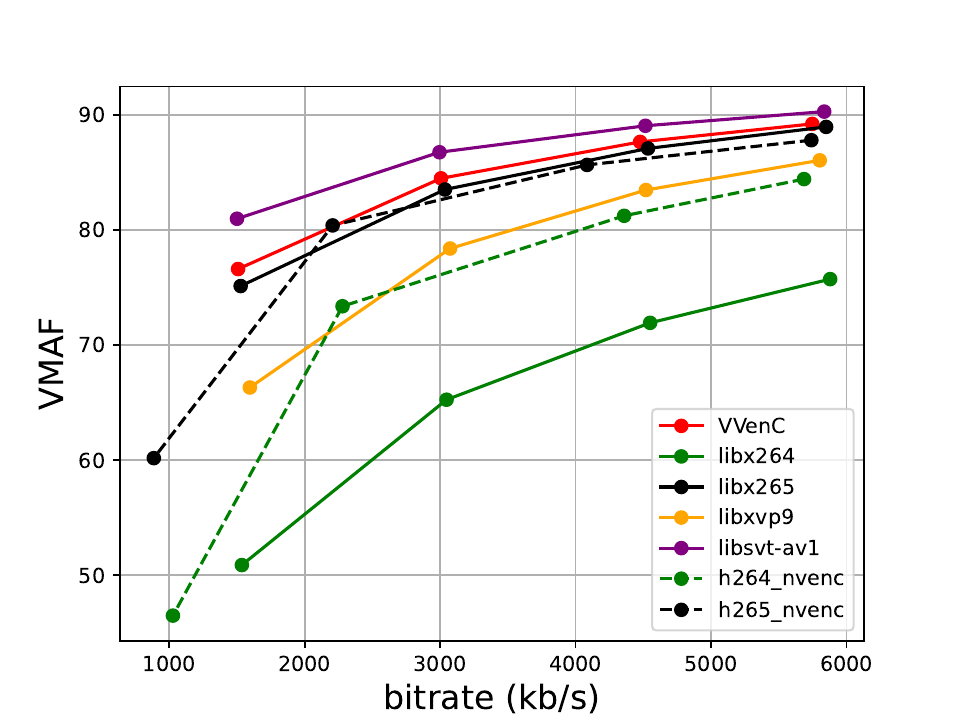}\label{fig-codec_res1c}}
    
    \caption{The average quality in S-PSNR (dB), SSIM and VMAF at different bit-rate for the seven considered encoders. \label{fig-codec_res1}}
    
\end{figure*}

Figures \ref{fig-codec_res1a}, \ref{fig-codec_res1b}, and \ref{fig-codec_res1c} provide the average quality performance of the studied software and hardware video encoders on the proposed dataset, utilizing three distinct quality metrics: S-\ac{psnr}, \ac{ssim}, and \ac{vmaf}, respectively. From the results, it is evident that the \ac{av1} software encoder achieves the highest quality in terms of S-\ac{psnr} and \ac{vmaf} across all four bitrates. Following closely is the \ac{vvenc} software encoder, which demonstrates competitive performance with \ac{av1}, particularly at high bitrates, based on the \ac{ssim} metric. On the other hand, the libx264 software encoder reports the lowest quality among the tested encoders. It is worth noting that the hardware design for the \ac{avc}/H.264 standard outperforms the libx264 software encoder significantly across all quality metrics and bitrates. Interestingly, the software implementation of the \ac{hevc}/H.265 standard exhibits slightly higher quality than its hardware implementation. This can be attributed to the increased complexity introduced by the new tools in the \ac{hevc}/H.265 standard, making the design of a hardware encoder for \ac{hevc} more challenging compared to the \ac{avc}/H.264 standard.

\begin{figure*}
\centering
    \subfigure[BD-PSNR]{\includegraphics[width=0.32\textwidth]{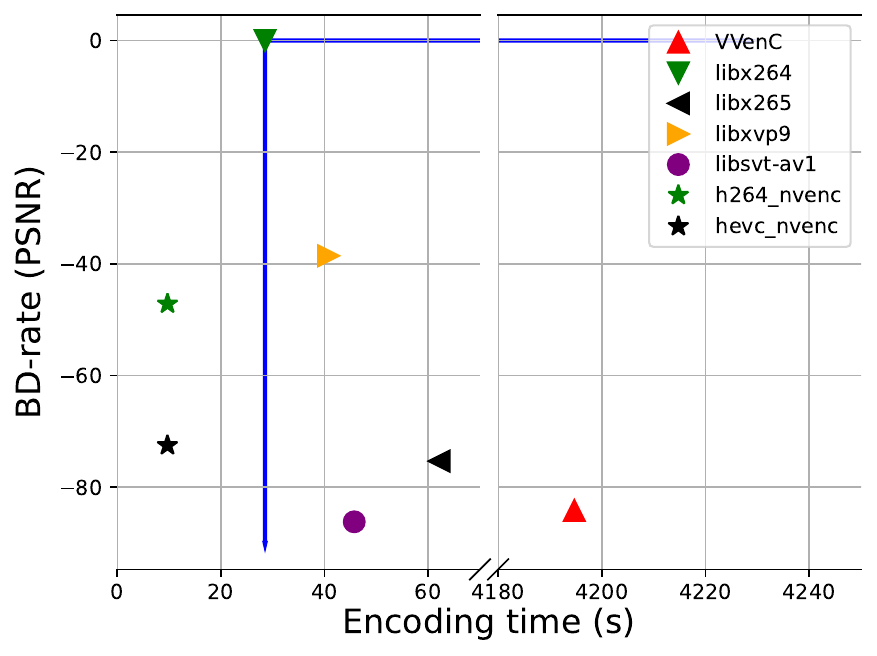}\label{fig-codec_resa}}    
    \subfigure[BD-SSIM]{\includegraphics[width=0.32\textwidth]{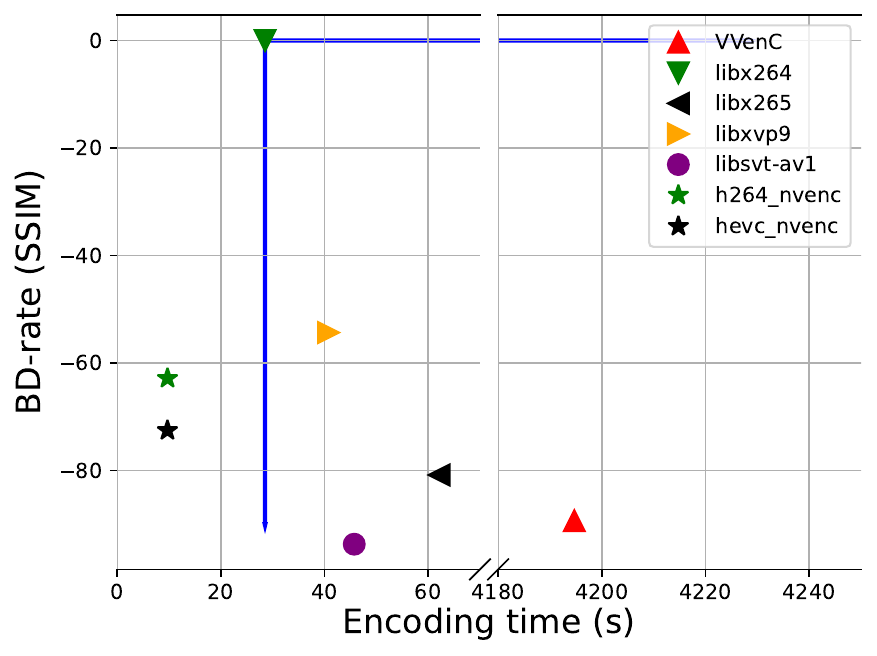}\label{fig-codec_resb}}    
    \subfigure[BD-VMAF] 
{\includegraphics[width=0.32\textwidth]{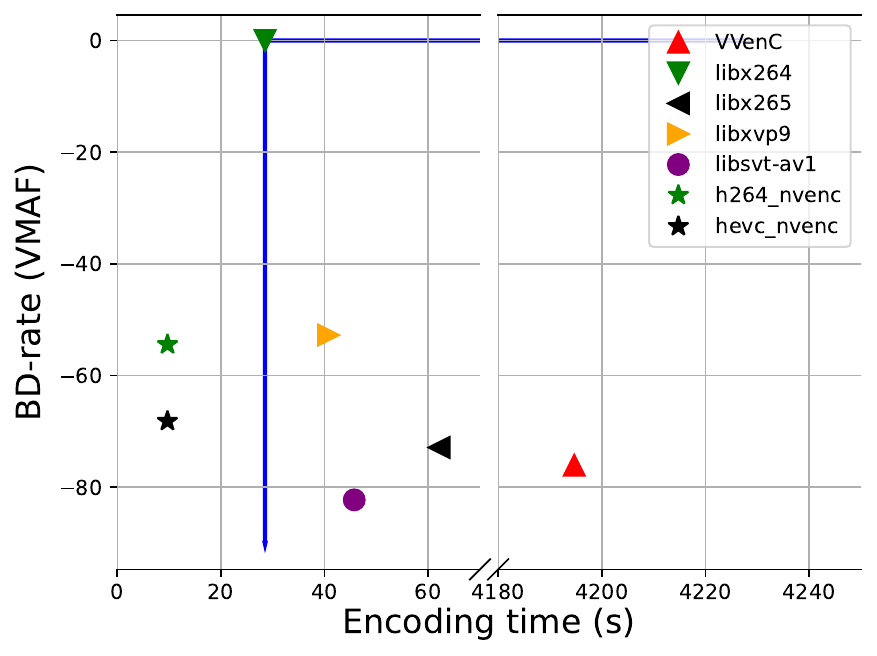} \label{fig-codec_resc}}   
    \caption{The \ac{bd-rate} performance in S-PSNR (dB), SSIM and VMAF versus encoding time for the seven considered encoders on 10 seconds video contents.  \label{fig-codec_res}}
\end{figure*}

The associated BD-rate results with respect to the \ac{avc}/H.264 software encoder for S-\ac{psnr}, \ac{ssim}, and \ac{vmaf} are depicted respectively in Figures \ref{fig-codec_resa}, \ref{fig-codec_resb} and \ref{fig-codec_resc}, plotted versus the encoding time. These figures reveal that the hardware encoders (h264\_nvenc and h265\_nvenc) and the \ac{av1} software encoder offer the best tradeoff between coding efficiency and encoding time. Notably, only the hardware encoders can achieve real-time encoding at 30 frames per second. To achieve real-time encoding, the \ac{av1}, \ac{avc}, and \acs{vp9} software encoders would require a powerful processor with multiple cores operating at a higher frequency. In contrast, the \ac{vvc}/H.266 software encoder (\ac{vvenc}) exhibits significantly longer encoding times, taking more than one hour to encode a 10-second video. The new coding tools introduced in the \ac{vvc}/H.266 standard have expanded the search space for rate-distortion optimizations, leading to increased encoding complexity. To enable real-time capability, advanced algorithmic optimizations, along with more efficient low-level optimizations, are necessary. Furthermore, the development of efficient hardware designs for the \ac{vvc}/H.266 standard becomes crucial, particularly for low-energy embedded devices, to achieve real-time encoding and benefit from its high coding efficiency and advanced features for \ac{odv} contents.

%We can notice from these figures that the encoders that achieve the best tradeoff between coding efficiency and encoding time are the two hardware encoders (i.e., h264\_nvenc, h265\_nvenc) and the \ac{av1} software encoder. More importantly, only the hardware encoders can achieve real-time encoding at 30 frames per second. The \ac{av1}, \ac{avc}, and VP9 software encoders would require an advanced processor with more cores running at a higher frequency to reach real-time encoding. In contrast, the  \ac{vvc}/H.266 software encoder \ac{vvenc} exhibits a very high encoding time, requiring more than one hour to encode a video of 10 seconds. The new coding tools introduced in the \ac{vvc}/H.266 standard have significantly expanded the search space explored by the rate-distortion optimizations of the encoder. Therefore, advanced algorithmic optimizations with more efficient low-level optimizations are required to reach real-time capability. Further, an efficient hardware design of the \ac{vvc}/H.266 standard is needed for real-time encoding, especially on low-energy embedded devices. 

%By comparatively studying the coding gain of different 360 \ac{uav} videos, it is observed that the rate and distortion efficiency is consistent among diverse kinds of contents. Therefore, Figure \ref{fig-codec_res1} shows the average performance in terms of S-PSNR, SSIM and VMAF, of all the dataset sequences. 


%From Figures \ref{fig-codec_res1b} and \ref{fig-codec_res1c} providein the coding performance in terms of S-PSNR and VMAF ,we can notice that the AV1 software encoder exhibits dominance across varied bit-rate targets, with a S-PSNR value exceeding the software AVC, the lowest-performing codec, by a substantial 46\%. However, the \ac{vvc} encoder outperforms the AV1 encoder in the realm of SSIM results for high bit-rate targets, achieving approximately 2\% greater S-PSNR value. Therefore, from a qualitative perspective, the \ac{vvc} and AV1 encoders, through their respective \textit{VVenC} and \textit{libsvtav1} software implementations, present optimal performance results when compared with other software and hardware implementations.

\begin{table}[t]
    \centering
    \caption{Video encoder \acs{sw}/\acs{hw} libraries}
        \resizebox{0.48\textwidth}{!}{%
    \begin{tabular}{llll}
    \toprule
        \textbf{Video codec standard} & \textbf{Software} & \textbf{Version} & \textbf{Hardware}   \\
         \midrule      
         \acs{avc}/H.264 & libx264 \cite{libx264} & 0.164.3106 & h264\_nvenc \cite{h264_nvenc} \\
         \midrule
          \acs{hevc}/H.265 & libx265 \cite{x265} & 3.5+1-f0c1022b6 & hevc\_nvenc \cite{hevc_nvenc}  \\
         \midrule
      \acs{vvc}/H.266 & \acs{vvenc} \cite{9455944} & 1.7.0  & - \\
         \midrule
         \acs{av1} & libsvtav1 \cite{libsvt_av1} & 1.4.1 & - \\
         \midrule
         \acs{vp9} & libvpx-vp9 \cite{libvpx_vp9} & 1.11.0-30-g888bafc78 & - \\
         \bottomrule
    \end{tabular}}
    \label{tab:codec}
\end{table}

%To further dive into the R-D efficiency interpretation of different codecs, we plot the BD results of different metrics: S-PSNR, SSIM and VMAF in Figure \ref{fig-codec_res1}: (a), (b) and (c), respectively. The results show that best performance trad-off between run-time and objective quality is clearly for hardware based encoders. While they maintain a good quality, the run time gap with other encoders is a significant advantage especially for live streaming. \textit{h264\_nvenc} and \textit{hevc\_nvenc} depicted by the green and black stars in the figures, are NVIDIA hardware implementations for AVC and HEVC, respectively. When compared against each other, \textit{hevc\_nvenc} has better results in terms of objective quality with around 40\% overhead compared \textit{h264\_nvenc}. Moreover, the two hardware encoders maintain closely the same run-time complexity.  Therefore, the hardware based implementation of HEVC has the best trade-off between rate and distortion.           

\subsection{Testbed for UAV 360$^\circ$ Video Streaming} 

%Our testbed setup involves a drone with a 360$^\circ$ camera as shown in Figure \ref{Scenario} (b) and an operator with a VR headset and a 5G connection, where they communicate through a central edge server.

%When the drone goes to high altitudes, there can be high interference that affects the quality of the video and the control signal in terms of latency. This interference is caused by the noise beams of the base stations that are down-tilted to serve ground-based users, not drones at high altitudes. As a result, the delay in the video and control signal can increase, making it more challenging for the operator to navigate the drone. Despite these potential challenges, our setup demonstrates the potential for using drones and VR headsets with 5G connectivity for various remote applications. However, it also highlights the need to address interference issues for better performance and reliability.
%A category 5 hurricane is approaching a coastal city, and the emergency response teams are bracing for the worst. The city's critical infrastructure, including power plants, water treatment plants, and communication towers, are all at risk of severe damage. In response, the team deploys a drone equipped with a 360$^\circ$ camera and 5G modem to capture high-resolution footage of the facilities in real-time such as shown in Figure \ref{Scenario} (a). The footage is transmitted to the command center for quick and accurate assessment of the situation \hl{This not the place, only scientific description is required. Go straight to "At high altitude, the drone encounters ..."}.

The proposed testbed comprises essential components, namely a \ac{uav} equipped with a 360-degree camera, a 5G modem, and an edge server. The 360-degree camera captures a comprehensive view of the surroundings, providing an immersive \ac{3dof} viewing experience that ensures no critical details are overlooked during acquisition. The 5G modem enables real-time transmission of high-resolution footage from the \ac{uav} to the edge server. Users can connect to the edge server through a \ac{hmd} at the command center, facilitating prompt decision-making by providing immediate access to live 4K 360-degree video footage.

\begin{figure}[t]
\centering
\subfigure[Field test]{\includegraphics[width=0.235\textwidth]{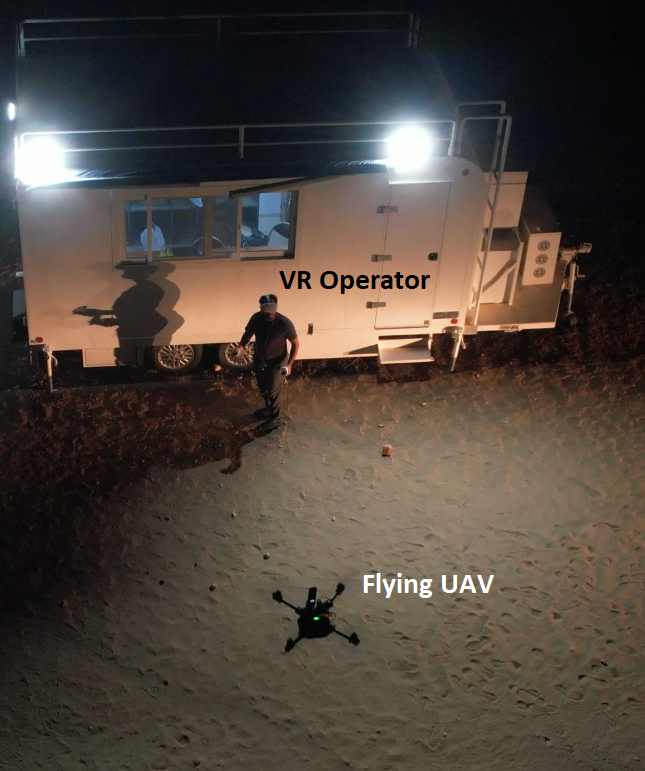}\label{Scenarioa}}
\subfigure[UAV]{\includegraphics[width=0.242\textwidth]{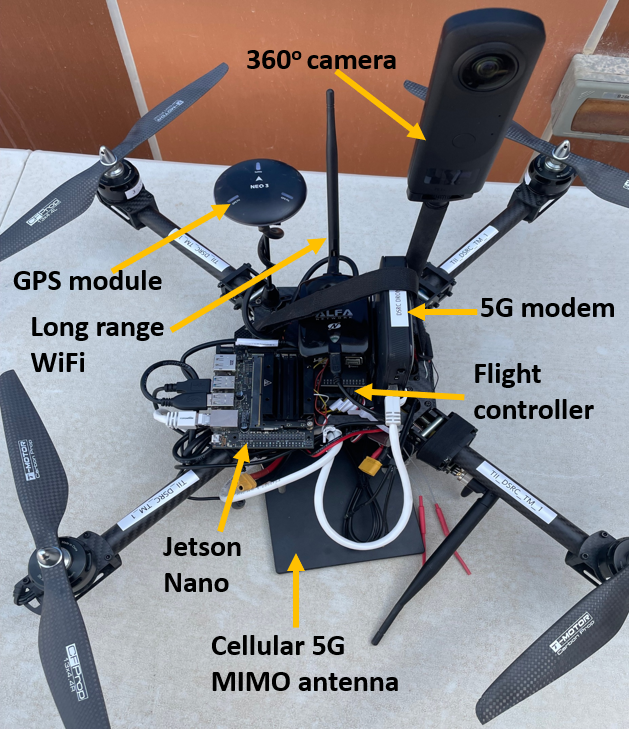}\label{Scenariob}}
        \caption{Illustration of the field test setting and the \acs{uav} configuration. \label{Scenario}}
        \vspace{-3mm}
\end{figure}

Figure \ref{Scenarioa} depicts the field tests conducted with a \ac{fpv} \ac{uav} operator controlling the \ac{uav} in a desert environment. The operator sends commands to the \ac{uav} through a central server located 100 km away from both the \ac{uav} and the operator. Both the \ac{uav} and the operator are connected to a consumer 5G network, as shown in Figure~\ref{Scenariob}, with specific settings outlined in Table \ref{table:testbed}. During the experiment, the operator flew the drone at a fixed position while varying the altitude. Simultaneously, the onboard computer of the \ac{uav} recorded information received from the 5G modem, including the Cell ID, throughput, and network latency from the \ac{uav} to the central server.

\begin{figure}[t]
\centering
    \subfigure[Handover vs. altitude.]{\includegraphics[width=0.234\textwidth]{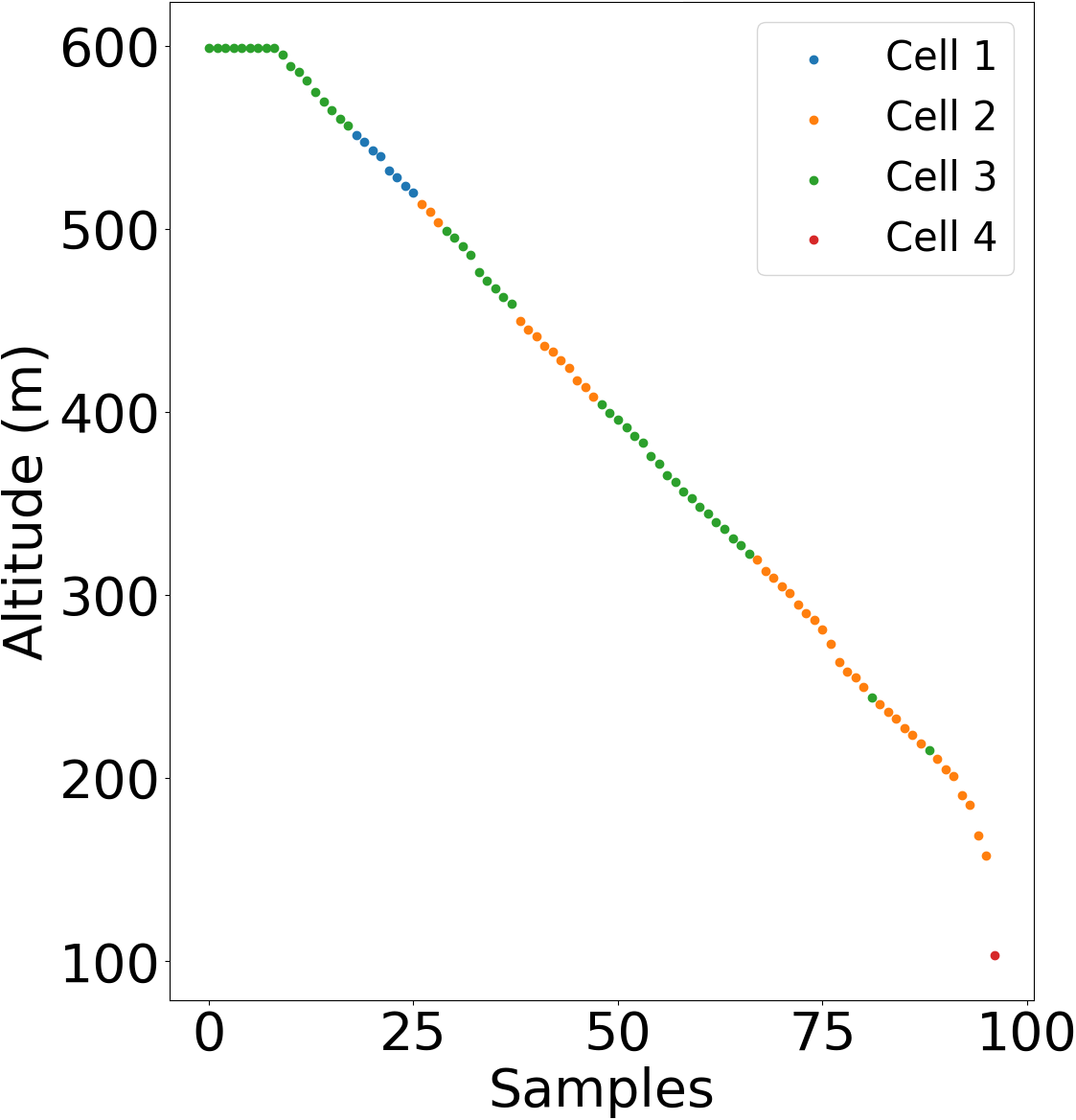} \label{results_uav_a}} \subfigure[Throughput vs. handover.]{\includegraphics[width=0.234\textwidth]{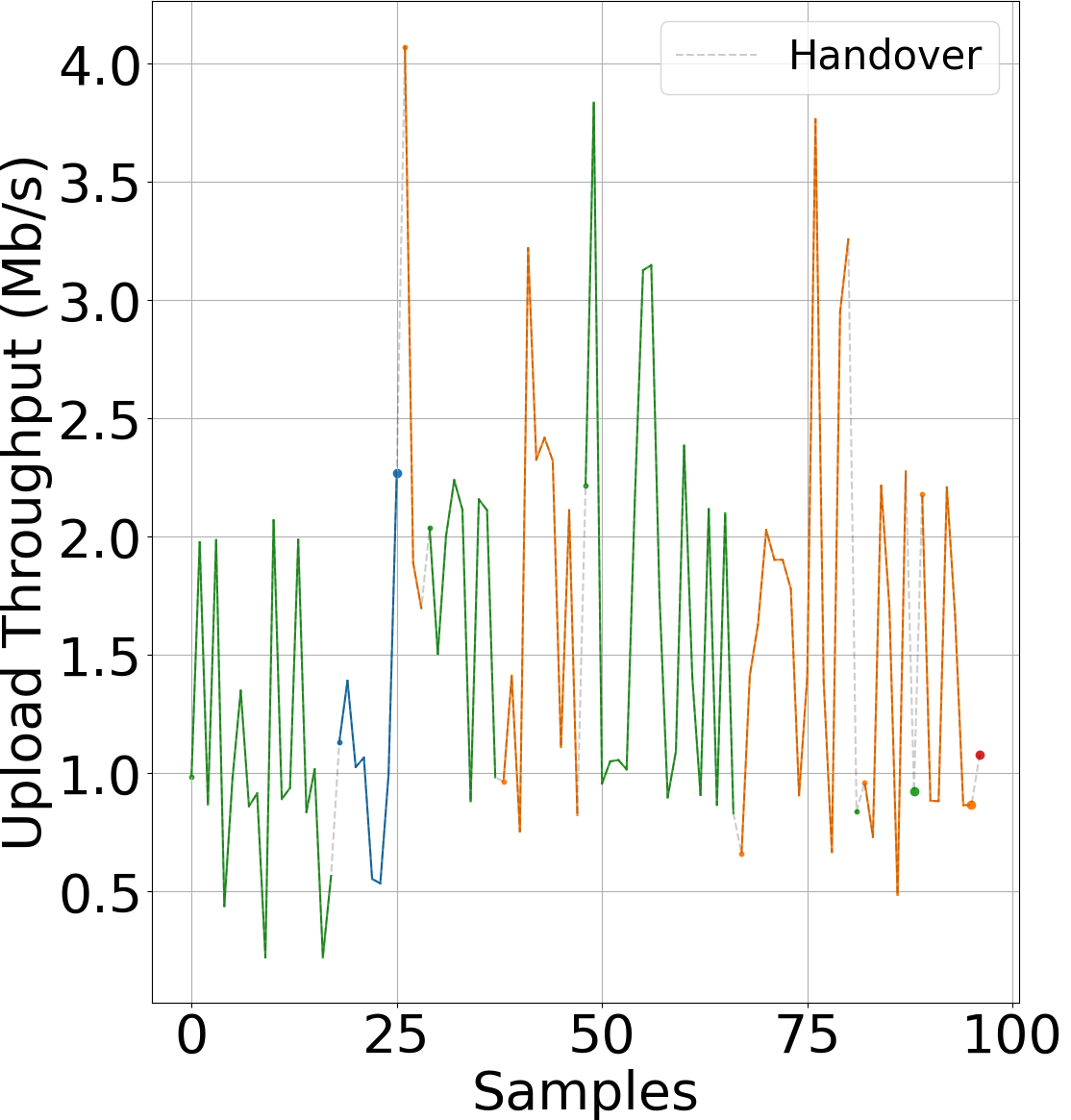} \label{results_uav_b}}
        \caption{Handovers and instantaneous throughput performance versus the drone altitude in vertical landing flying conditions. The average throughput values of the cells in green, orange, blue, and red are 14.55 Mbps, 17.19 Mbps, 11.21 Mbps, and 10.79 Mbps, respectively.  \label{results_uav}} \vspace{-3mm}
\end{figure}

Figures \ref{results_uav_a} and \ref{results_uav_b} provide insights into the flying conditions for handovers and the instantaneous throughput as a function of altitude in the scenario of vertical landing of the drone. In Figure \ref{results_uav_a}, it can be observed that the drone experienced a total of ten handovers, utilizing the four available base stations that cover the flying area. Figure \ref{results_uav_b} shows that most of the handovers resulted in improved instantaneous throughput. However, the throughput exhibited significant fluctuations due to wireless communication instability and interference.

At higher altitudes, the drone encounters interference from base stations primarily designed for ground-based users. This interference introduces latency and quality degradation in the transmission of video and control signals, thereby posing challenges for effective drone navigation by the operator. Our real field tests showcase the control of \ac{uav} through 5G using a \ac{vr} headset and 360-degree video feedback at altitudes of up to 600 meters. These tests shed light on the potential challenges associated with interfering base stations and suboptimal handover conditions in \ac{vr}-based \ac{uav} control.

%Figures \ref{results_uav_a} and \ref{results_uav_b} present the flying conditions for handovers and instantaneous throughput versus altitude in the scenario of vertical landing of the drone. In Figure~\ref{results_uav_a}, we observe that the drone underwent a total of ten handovers, utilizing the available four base stations that cover the flying area. Figure \ref{results_uav_b} indicates that most of the handovers resulted in improved instantaneous throughput. However, the throughput experienced significant fluctuations due to wireless communication instability and interference.

%At higher altitudes, the drone faces interference from base stations designed for ground-based users. This interference leads to latency and quality degradation of the video and control signals, posing challenges for effective drone navigation by the operator. Our real field tests demonstrate UAV control through 5G with a VR headset and 360$^\circ$ video feedback at altitudes up to 600m. These tests highlight the potential challenges posed by interfering base stations and suboptimal handover conditions in \ac{vr}-based \ac{uav} control.

\begin{table}[t]    \caption{Configuration of the 360$^\circ$ video streaming over \acs{uav} testbed.}   
\begin{tabular}{p{0.5\linewidth}p{0.4\linewidth}}
\hline
Parameter  &  Value \\
\hline
5G Max(upload/download) & 50 Mbps/100 Mbps \\
Server \acs{cpu} & 8 cors @ 2.5 GHz\\
Server memory & 16 Gb \\
Distance \acs{uav} to Server & 500m \\
Distance \ac{vr} \ac{hmd}/\acs{uav} to Server & 100km \\
\acs{uav} flight speed during tests & 25Km/h \\
\acs{uav}'s onboard computer & Jetson nano\\
\acs{uav}'s weight &  2.5 Kg \\
360-degree camera  & Ricoh Theta Z1\\
\hline
\end{tabular}
\label{table:testbed}
\end{table}

\section{Open Challenges}
\label{sec:chal}  

\subsection{Adaptive Low-latency 360$^\circ$ Video Streaming}
From Figure~\ref{results_uav_b}, it is evident that \ac{uav} communication, particularly at high altitudes and during mobility, is susceptible to significant throughput variation. This inherent issue raises fundamental concerns regarding the attainment of high \ac{qoe} with superior video quality and minimal \ac{g2g} latency in real-time 360-degree video streaming. To address these challenges, several open research directions can be pursued. Firstly, leveraging the latest video coding standards and efficient hardware encoders, such as hevc\_nvenc, can substantially enhance perceived video quality. The hevc\_nvenc encoder enables real-time encoding with low energy consumption, harnessing the coding efficiency promised by the advanced video coding standard, \ac{hevc}/H.265. This, in turn, extends the \ac{uav}'s battery life. At the cloud level, more efficient software encoders like SVT-\ac{av1} can be utilized for video transcoding, leveraging the available cloud resources. Furthermore, advanced optimization techniques, such as \ac{fov} prediction, can be employed to allocate higher quality to the viewing viewport, resulting in improved bandwidth utilization and perceived video quality for end-users. 

Secondly, ensuring fast and accurate adaptation of the video bitrate to channel throughput variations is crucial to prevent buffering at both the transmitter and receiver sides, thereby minimizing \ac{g2g} latency. In this regard, leveraging information from the physical layer as well as the \ac{uav} status, position and its environment can be valuable for predicting throughput variations and facilitating proactive encoder adaptation. Additionally, jointly considering other source video parameters, such as spatial resolution, temporal frame rate, and projection, in the rate control mechanism can further minimize \ac{g2g} latency and maximize perceived quality. Advanced machine learning techniques, including deep reinforcement learning, have demonstrated potential in learning real-time prediction of pre-processing and encoder parameters to achieve the target bitrate while maximizing perceived quality \cite{huang2021deep, mandhane2022muzero}.

Finally, these research directions pave the way for addressing open challenges in UAV-based 360-degree video streaming, leading to improved \ac{qoe}, minimized latency, and enhanced video quality.

\subsection{Cooperative Aerial Video Streaming}
Cooperative immersive video streaming, exemplified by Intel's Trueview \cite{IntelTrueview}, has the potential to enable a truly immersive viewing experience \cite{Wang_ICDCS_2017}. This approach allows users to independently select their preferred viewing angle by streaming from multiple cameras or sources, leveraging spatial diversity in terms of viewing angle, content, or geographic location. Moreover, employing multiple \acp{uav} to capture aerial views can enhance the immersive experience with \ac{6dof} capabilities \cite{Meng_ACMMulti_2015, Wang_MILCOM2022}. However, developing a multi-\ac{uav} cooperative immersive video streaming system entails addressing a unique set of challenges in joint communication, computation, and control design. Streaming a scene captured by multiple \acp{uav} requires effective coordination among the \acp{uav} to ensure comprehensive scene coverage without compromising \ac{qoe} while minimizing network bandwidth usage. Additionally, capturing more dynamic events, such as sports or moving ground targets \cite{Wang_ICDCS_2017, Wang_MILCOM2022}, necessitates accurate motion prediction, such as player or target movement, which, in turn, relies on coordinated trajectory planning and 3D placement of all \acp{uav}. Furthermore, the trajectory and placement of \acp{uav} must also consider their battery levels, in addition to \ac{qoe} considerations.

In multi-\ac{uav} applications, the individual \ac{uav} can collaboratively and cohesively capture videos, which are then synthesized into a panoramic video. However, streaming videos from all \acp{uav} simultaneously poses a significant resource burden. To address this challenge, bandwidth-saving streaming techniques can be employed by leveraging user attention information \cite{WanDuLiSon21}. Specifically, \acp{uav} whose videos are deemed unnoticed by users can remain idle during transmission. However, we argue that instead of staying idle, these \acp{uav} can contribute to real-time video streaming, thus enhancing communication efficiency and throughput further. For instance, the \ac{uav} swarm can collectively form a virtual \ac{mimo} system \cite{QiaWanJin22}. Nonetheless, this type of \ac{mimo} system exhibits distinct wireless channel characteristics. Considering the unique channel model and the requirements for throughput and latency, thus designing cooperative aerial video streaming for real-time and interactive panoramic videos poses considerable challenges.

Addressing these challenges in cooperative aerial video streaming requires innovative solutions that account for coordination, resource optimization, wireless channel characteristics, throughput, and latency requirements. Meeting these objectives will contribute to the development of robust and efficient systems for real-time and interactive panoramic video streaming.

\subsection{\acs{qoe}-Aware Control and Communication of \acp{uav}}
Ensuring high-quality user experience in 360$^\circ$ video streaming systems is primarily affected by stall time resulting from low transmission rates and overall video quality perceived by users \cite{Anwar_IEEEAccess_2020}. Experiencing a stall time longer than the tolerance level can lead to \ac{vr} sickness, making it crucial to achieve high data rates and low-latency transmission for enhancing \ac{qoe} \cite{Teng_TWC_Jan2022}. Addressing these challenges requires leveraging video saliency to predict users' \ac{fov} and employing multicast transmission techniques based on users' locations and \ac{fov} correlations, as grouping and multicasting can improve network throughput and \ac{qoe} \cite{Huang_ArXiv2018, Cristina_TC_Apr2020}. Additionally, adapting the encoding bitrate of tiles based on channel quality, available resources, content quality, and inaccurate \ac{fov} prediction can further enhance \ac{qoe} \cite{Teng_TWC_Jan2022}.

In the context of \ac{vr} streaming from aerial users, such as drones, additional challenges arise due to their dynamic topology and limited energy resources \cite{Ding_Globecom_2019, Shirey_ISQoS_2021}. The channel quality and network throughput of aerial users are also influenced by their flight trajectory, necessitating the joint design of \ac{qoe}-aware resource allocation and drone route selection mechanisms \cite{Colonnese_IEEESAM_2018, Tang_TMC_July2022}. Moreover, the limited onboard energy availability of drones requires judicious resource allocation strategies \cite{Yang_IEEEAccess_2019}. Furthermore, in a multi-\ac{uav}-based streaming system for 360$^\circ$ videos \cite{HeXieTia19}, additional challenges arise in terms of resource allocation among the \acp{uav}. Each \ac{uav} can independently adjust its encoding bitrate and position \cite{Liyana_TVT_May2022}, while simultaneously competing for resources with other \acp{uav} in the swarm. Given these significant challenges, designing a \ac{uav}-based 360$^\circ$ streaming system necessitates a thorough study of joint communication and control design for these systems.

In critical missions involving the teleoperation of \acp{uav}, such as fire disaster monitoring \cite{Nihei22} and suspicious vehicle tracking \cite{LiuZhuDenGuaWanLuoLinZha19}, the quality of service relies on the interplay between control command delivery and video data transmission. The latency experienced in one link can impact the latency budget in the other link. Moreover, unreliable control command communication can influence the \ac{uav}'s reaction and view angle, resulting in undesired information for the remote operator. Therefore, the entanglement and mutual influence between control command delivery and real-time \acp{uav} video transmission require dedicated consideration in the design process.

\begin{figure}[th]
\centering
\includegraphics[width=8cm]{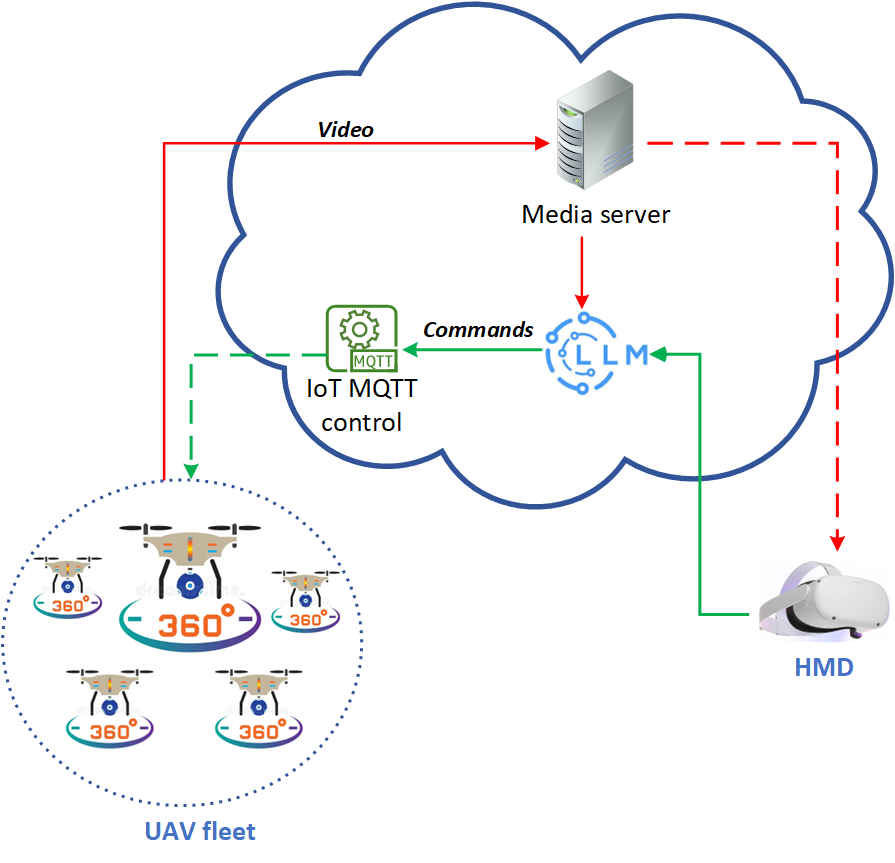}
\caption{Use case scenario for \acp{llm} control commands for \acs{uav} with 360$^\circ$ camera.}
\vspace{-3mm}
\end{figure}

\subsection{Tailored Design of \acs{uav} Communication for Video Streaming}
%Various features of real-time video streaming have been considered in the theoretical design of \ac{uav} communication systems. Many theoretical solutions and algorithms were proposed for real-time \ac{uav} video streaming by incorporating some distinctive characteristics of video streaming and air-to-ground channel as well as the communication requirements. In testbed implementation, video streaming can dynamically adapt to wireless conditions, but the communication algorithm is not tailored to video streaming. In other words, due to the simplicity, the existing protocols are utilized to evaluate the real-time \ac{uav} video streaming performance in the real world. To have more insights into enabling immersive services by 6G networks, implementing the wireless protocols tailored to video streaming demands more effort. Nevertheless, the current wireless system merely considers the video capture rate or bitrate.

%Intuitively, video streaming-tailored wireless systems can be achieved by involving more characteristics in video encoding in the wireless system design.
%For instance, the video content correlation among frames can be considered to enhance video encoding efficiency. Furthermore, in semantic communication \cite{CalBar21}, the main goal is to effectively convey the semantic meaning of information instead of reliably delivering each digital bit. Hence, analyzing the video content correlation and the content meaning, semantic communication provides a promising framework to realize the video streaming-tailored UAV communication. 
The design and optimization of \ac{uav} wireless communication systems for real-time video streaming pose several open challenges. While theoretical solutions and algorithms have been proposed, many of them focus on incorporating specific characteristics of video streaming and air-to-ground channels, as well as addressing communication requirements. However, in practical testbed implementations, the existing protocols are often used without tailored optimization for video streaming, resulting in limited performance evaluation in real-world scenarios. To enable more immersive services in the context of 6G networks, it is essential to dedicate effort towards implementing wireless protocols specifically tailored to the demands of video streaming. Currently, the wireless system primarily considers factors such as video capture rate or bitrate, neglecting other important aspects.

One approach to achieve video streaming-tailored wireless systems is to incorporate additional characteristics of video encoding in the design. For example, considering the correlation among frames in video content can enhance video encoding efficiency. Additionally, adopting a semantic communication approach, as explored in \cite{CalBar21}, where the focus is on effectively conveying the semantic meaning of information rather than solely delivering digital bits, shows promise in realizing \ac{uav} communication systems tailored for video streaming. Addressing these challenges and exploring novel techniques that take into account video content correlation, semantic communication, and other relevant factors will be crucial in developing efficient and optimized wireless systems for \ac{uav} video streaming applications.

%\cite{Ge_TMM_Oct2017, Chakareski_MMSP_2020, Mahzari_ISM_2020, Yixuan_ICME_2020, Yang_JIoT_sep2022} 
\subsection{\acs{llm} for Immersive Video Streaming}
%The rapid advancement in \ac{nlp} has paved the way for development of \acp{llm}, such as BERT \cite{47751}, \acs{gpt}-3/\acs{gpt}-4, and FALCON. These models are explored in many ways from conversational, medical, Telecom, and to robotic applications. These models can be explored to enhance the key performance when streaming 360-degree video from one or multiple \acp{uav}. For instance, the end used can prompt the \acp{llm} with the task to complete along the description of the environment. The \acp{llm} can then provide the command to follow by the \acp{uav} to complete the tasks while minimizing their energy consumption. 
The rapid advancement in \ac{nlp} has paved the way for the development of \acp{llm} like BERT \cite{47751}, \acs{gpt}-3/\acs{gpt}-4, and FALCON. These versatile models push the state-of-the-art on many down-stream tasks, finding applications in various domains, including conversation, medicine, telecommunications \cite{bariah2023large}, and robotics \cite{vemprala2023chatgpt}.

In the context of streaming 360-degree video from one or multiple \acp{uav}, exploring these \acp{llm} can greatly enhance performance. For example, end users can provide task prompts to the \ac{llm} along with descriptions of the environment captured by the 360-degree camera. The \ac{llm} can then generate commands for the \acp{uav} to carry out the tasks successfully while minimizing their energy consumption and avoiding obstacles. In particular, the description of the surrounding environment from the 360-degree camera can be performed by the end used or automatically by exploiting vision-language models, such as SimVLM~\cite{wang2022simvlm}, Flamingo~\cite{alayrac2022flamingo}, or BLIP-2~\cite{li2023blip2}.

\section{Conclusion}
\label{sec:con}

In this paper, we have investigated  omnidirectional video streaming over \acp{uav}, focusing on the benefits and challenges associated with live 360-degree video streaming. By enabling immersive viewing with up to \ac{6dof}, this technology enhances the \ac{qoe} for various applications such as surveillance, autonomous driving, healthcare, and education. However, 360-degree video streaming poses challenges in terms of high bandwidth and computing requirements, while the \ac{uav} wireless channel exhibits interference and instability, leading to significant bandwidth variations. To overcome these challenges, we first reviewed the key components of 360-degree video streaming over wireless channels and highlighted the technology used to achieve low-latency end-to-end streaming. Additionally, we introduced a new dataset consisting of ten 360-degree videos captured by \ac{uav} in various flying conditions, enabling us to evaluate the coding efficiency and complexity of different \ac{sw} and \ac{hw} video encoders. Through our experiments, we found that only \acs{hw} \acs{avc}/H.264 and \acs{hevc}/H.265 encoders achieved real-time encoding, making them suitable for \acs{uav} with limited computing and energy resources. Furthermore, the \acs{sw} AV1 encoder demonstrated the best tradeoff between coding performance and complexity and thus be utilized for efficient video transcoding on more powerful devices in the cloud. Moreover, we presented a real testbed of 360-degree video streaming over a drone with 5G communication, illustrating the significant fluctuations in the wireless channel due to interference and multiple handovers. Finally, we discussed open challenges and proposed future research directions to enhance the key performance metrics of live immersive video streaming over \acp{uav}. 

Overall, this paper provides valuable insights into the field of omnidirectional video streaming over \acp{uav} and contributes to the understanding of how to improve the \ac{qoe} in this context. The findings and recommendations presented here pave the way for further advancements and innovations in the area of live immersive video streaming over \acp{uav}, ultimately benefiting a wide range of applications and industries.

\bibliographystyle{IEEEtran}
\bibliography{biblio}

\begin{thebibliography}{100}
\providecommand{\url}[1]{#1}
\csname url@samestyle\endcsname
\providecommand{\newblock}{\relax}
\providecommand{\bibinfo}[2]{#2}
\providecommand{\BIBentrySTDinterwordspacing}{\spaceskip=0pt\relax}
\providecommand{\BIBentryALTinterwordstretchfactor}{4}
\providecommand{\BIBentryALTinterwordspacing}{\spaceskip=\fontdimen2\font plus
\BIBentryALTinterwordstretchfactor\fontdimen3\font minus
  \fontdimen4\font\relax}
\providecommand{\BIBforeignlanguage}[2]{{%
\expandafter\ifx\csname l@#1\endcsname\relax
\typeout{** WARNING: IEEEtran.bst: No hyphenation pattern has been}%
\typeout{** loaded for the language `#1'. Using the pattern for}%
\typeout{** the default language instead.}%
\else
\language=\csname l@#1\endcsname
\fi
#2}}
\providecommand{\BIBdecl}{\relax}
\BIBdecl

\bibitem{10089176}
J.~van~der Hooft, H.~Amirpour, M.~T. Vega, Y.~Sanchez, R.~Schatz, T.~Schierl,
  and C.~Timmerer, ``A tutorial on immersive video delivery: From
  omnidirectional video to holography,'' \emph{IEEE Communications Surveys \&
  Tutorials}, vol.~25, no.~2, pp. 1336--1375, 2023.

\bibitem{8660516}
M.~Mozaffari, W.~Saad, M.~Bennis, Y.-H. Nam, and M.~Debbah, ``A tutorial on
  {UAVs} for wireless networks: Applications, challenges, and open problems,''
  \emph{IEEE Commun. Surveys Tuts.}, vol.~21, no.~3, pp. 2334--2360, 3rd Quart.
  2019.

\bibitem{Wu_JSAC_Oct2021}
Q.~Wu, J.~Xu, Y.~Zeng, D.~W.~K. Ng, N.~Al-Dhahir, R.~Schober, and A.~L.
  Swindlehurst, ``A comprehensive overview on {5G}-and-beyond networks with
  {UAVs}: From communications to sensing and intelligence,'' \emph{IEEE J. Sel.
  Areas Commun.}, vol.~39, no.~10, pp. 2912--2945, Dec. 2021.

\bibitem{Hayat_IEEECommSurvey_Dec2016}
S.~Hayat, E.~Yanmaz, and R.~Muzaffar, ``Survey on unmanned aerial vehicle
  networks for civil applications: A communications viewpoint,'' \emph{IEEE
  Commun. Surveys Tuts.}, vol.~18, no.~4, pp. 2624--2661, 4th Quart. 2016.

\bibitem{Baltaci_CommSurvey_April2021}
A.~Baltaci, E.~Dinc, M.~Ozger, A.~Alabbasi, C.~Cavdar, and D.~Schupke, ``A
  survey of wireless networks for future aerial communications {(FACOM)},''
  \emph{IEEE Commun. Surveys Tuts.}, vol.~23, no.~4, pp. 2833--2884, 4th Quart.
  2021.

\bibitem{Mishra_IEEECommstandardMag_May2020}
D.~Mishra, A.~M. Vegni, V.~Loscrí, and E.~Natalizio, ``Drone networking in the
  {6G} era: A technology overview,'' \emph{IEEE Commun. Standards Mag.},
  vol.~5, no.~4, pp. 88--95, Dec. 2021.

\bibitem{Yan_IEEE_acess_Jul2019}
C.~Yan, L.~Fu, J.~Zhang, and J.~Wang, ``A comprehensive survey on {UAV}
  communication channel modeling,'' \emph{IEEE Access}, vol.~7, pp.
  107\,769--107\,792, 2019.

\bibitem{Zhenyu_IEEEComm_survey_Jan2022}
Z.~Xiao, L.~Zhu, Y.~Liu, P.~Yi, R.~Zhang, X.-G. Xia, and R.~Schober, ``A survey
  on millimeter-wave beamforming enabled {UAV} communications and networking,''
  \emph{IEEE Commun. Surveys Tuts.}, vol.~24, no.~1, pp. 557--610, 1st Quart.
  2022.

\bibitem{Yongs_IEEEProc_Dec2019}
Y.~Zeng, Q.~Wu, and R.~Zhang, ``Accessing from the sky: A tutorial on {UAV}
  communications for {5G} and beyond,'' \emph{Proc. IEEE}, vol. 107, no.~12,
  pp. 2327--2375, Dec. 2019.

\bibitem{Elmokadem_Sensors_2021}
T.~Elmokadem and A.~V. Savkin, ``Towards fully autonomous {UAVs}: A survey,''
  \emph{Sensors}, vol.~21, no.~18, pp. 1--39, Sep. 2021.

\bibitem{Fotouhi_IEEEComSurvey_Dec2019}
A.~Fotouhi, H.~Qiang, M.~Ding, M.~Hassan, L.~G. Giordano, A.~Garcia-Rodriguez,
  and J.~Yuan, ``Survey on {UAV} cellular communications: Practical aspects,
  standardization advancements, regulation, and security challenges,''
  \emph{IEEE Commun. Surveys Tuts.}, vol.~21, no.~4, pp. 3417--3442, 4th Quart.
  2019.

\bibitem{Marojevic_VT_Magazine_Feb2020}
V.~Marojevic, I.~Guvenc, R.~Dutta, M.~L. Sichitiu, and B.~A. Floyd, ``Advanced
  wireless for unmanned aerial systems: {5G} standardization, research
  challenges, and {AERPAW} architecture,'' \emph{IEEE Veh. Technol. Mag.},
  vol.~15, no.~2, pp. 22--30, Jun. 2020.

\bibitem{Abdalla_IEEE_commstandardMag_May2021}
A.~S. Abdalla and V.~Marojevic, ``Communications standards for unmanned
  aircraft systems: The {3GPP} perspective and research drivers,'' \emph{IEEE
  Commun. Standards Mag.}, vol.~5, no.~1, pp. 70--77, Mar. 2021.

\bibitem{Wang_IEEECommSurvey_Feb2020}
H.~Wang, H.~Zhao, J.~Zhang, D.~Ma, J.~Li, and J.~Wei, ``Survey on unmanned
  aerial vehicle networks: A cyber physical system perspective,'' \emph{IEEE
  Commun. Surveys Tuts.}, vol.~22, no.~2, pp. 1027--1070, 2nd Quart. 2020.

\bibitem{Yaqoob_IEEEComSurvey_2020}
A.~Yaqoob, T.~Bi, and G.-M. Muntean, ``A survey on adaptive 360$^\circ$ video
  streaming: Solutions, challenges and opportunities,'' \emph{IEEE Commun.
  Surveys Tuts.s}, vol.~22, no.~4, pp. 2801--2838, 4th Quart. 2020.

\bibitem{Dongbiao_ICETE_2018}
D.~He, C.~Westphal, and J.~J. Garcia-Luna-Aceves, ``Network support for {AR/VR}
  and immersive video application: A survey,'' in \emph{Proc. 15th Int. Joint
  Conf. e-Business Telecommun.}, Jul. 2018, pp. 359--369.

\bibitem{NGUYEN2023103564}
T.-V. Nguyen, N.~P. Nguyen, C.~Kim, and N.-N. Dao, ``Intelligent aerial video
  streaming: Achievements and challenges,'' \emph{J. Netw. Comput. Appl.}, vol.
  211, p. 103564, Feb. 2023.

\bibitem{Zink_IEEEProc_2019}
M.~Zink, R.~Sitaraman, and K.~Nahrstedt, ``Scalable 360° video stream
  delivery: Challenges, solutions, and opportunities,'' \emph{Proc. IEEE}, vol.
  107, no.~4, pp. 639--650, Apr. 2019.

\bibitem{9633229}
I.-C. Lo, K.-T. Shih, and H.~H. Chen, ``Efficient and accurate stitching for
  360° dual-fisheye images and videos,'' \emph{IEEE Trans. Image Process.},
  vol.~31, pp. 251--262, 2022.

\bibitem{8902161}
Y.~Ye, J.~M. Boyce, and P.~Hanhart, ``Omnidirectional 360° video coding
  technology in responses to the joint call for proposals on video compression
  with capability beyond {HEVC},'' \emph{IEEE Transactions on Circuits and
  Systems for Video Technology}, vol.~30, no.~5, pp. 1241--1252, 2020.

\bibitem{360meta}
\BIBentryALTinterwordspacing
E.~Kuzyakov and D.~Pio. (2016, Jan.) Next-generation video encoding techniques
  for 360 video and {VR}. Accessed on 1-July-2023. [Online]. Available:
  \url{https://engineering.fb.com/2016/01/21/virtual-reality/next-generation-video-encoding-techniques-for-360-video-and-vr/}
\BIBentrySTDinterwordspacing

\bibitem{1218189}
T.~Wiegand, G.~Sullivan, G.~Bjontegaard, and A.~Luthra, ``Overview of the
  {H.264/AVC} video coding standard,'' \emph{IEEE Trans. Circuits Syst. Video
  Technol.}, vol.~13, no.~7, pp. 560--576, Jul. 2003.

\bibitem{6316136}
G.~J. Sullivan, J.-R. Ohm, W.-J. Han, and T.~Wiegand, ``Overview of the high
  efficiency video coding {(HEVC)} standard,'' \emph{IEEE Trans. Circuits Syst.
  Video Technol.}, vol.~22, no.~12, pp. 1649--1668, Dec. 2012.

\bibitem{9503377}
B.~Bross, Y.-K. Wang, Y.~Ye, S.~Liu, J.~Chen, G.~J. Sullivan, and J.-R. Ohm,
  ``Overview of the versatile video coding {(VVC)} standard and its
  applications,'' \emph{IEEE Trans. Circuits Syst. Video Technol.}, vol.~31,
  no.~10, pp. 3736--3764, Dec. 2021.

\bibitem{6547985}
K.~Misra, A.~Segall, M.~Horowitz, S.~Xu, A.~Fuldseth, and M.~Zhou, ``An
  overview of tiles in {HEVC},'' \emph{IEEE J. Sel. Topics Signal Process.},
  vol.~7, no.~6, pp. 969--977, Dec. 2013.

\bibitem{10.1145/3210445.3210455}
J.~Son, D.~Jang, and E.-S. Ryu, ``Implementing motion-constrained tile and
  viewport extraction for {VR} streaming,'' in \emph{Proc. 28th ACM SIGMM
  Workshop Netw. Oper. Syst. Support Digit. Audio Video}, Jun. 2018, pp.
  61--66.

\bibitem{9171593}
N.~Mahmoudian~Bidgoli, T.~Maugey, and A.~Roumy, ``Fine granularity access in
  interactive compression of 360-degree images based on rate-adaptive channel
  codes,'' \emph{IEEE Trans. Multimedia}, vol.~23, pp. 2868--2882, 2021.

\bibitem{8926340}
Y.~Zhou, L.~Tian, C.~Zhu, X.~Jin, and Y.~Sun, ``Video coding optimization for
  virtual reality 360-degree source,'' \emph{IEEE J. Sel. Topics Signal
  Process.}, vol.~14, no.~1, pp. 118--129, Jan. 2020.

\bibitem{9000878}
Y.-C. Su and K.~Grauman, ``Learning compressible 360$^{\circ }$ video
  isomers,'' \emph{IEEE Trans. Pattern Anal. Mach. Intell.}, vol.~43, no.~8,
  pp. 2697--2709, Aug. 2021.

\bibitem{minnen2018joint}
D.~Minnen, J.~Ball{\'e}, and G.~D. Toderici, ``Joint autoregressive and
  hierarchical priors for learned image compression,'' \emph{Adv. Neural Inf.
  Process. Syst.}, vol.~31, 2018.

\bibitem{9751563}
J.~Cao, X.~Su, B.~Finley, A.~Pauanne, M.~Ammar, and P.~Hui, ``Evaluating
  multimedia protocols on {5G} edge for mobile augmented reality,'' in
  \emph{Proc. 17th Int. Conf. Mobility, Sens. Netw.}, Dec. 2021, pp. 199--206.

\bibitem{9380215}
M.~M. Hannuksela and Y.-K. Wang, ``An overview of omnidirectional media format
  {(OMAF)},'' \emph{Proc. IEEE}, vol. 109, no.~9, pp. 1590--1606, Sep. 2021.

\bibitem{7160422}
B.~Sredojev, D.~Samardzija, and D.~Posarac, ``{WebRTC} technology overview and
  signaling solution design and implementation,'' in \emph{Proc. 38th Int.
  Convention Inf. Commun. Technol., Electron. Microelectron.}, May 2015, pp.
  1006--1009.

\bibitem{10.1145/3339825.3393576}
K.~K. Sreedhar, I.~D.~D. Curcio, A.~Hourunranta, and M.~Lepist\"{o},
  ``Immersive media experience with {MPEG OMAF} multi-viewpoints and
  overlays,'' in \emph{Proc. 11th ACM Multimedia Syst. Conf.}, May 2020, pp.
  333--336.

\bibitem{TileMedia1}
\BIBentryALTinterwordspacing
TileMedia. (2022, Jan.) {How ClearVR Drives and Leverages Standards}. Accessed
  on 1-July-2023. [Online]. Available: \url{https://www.tiledmedia.com/
  how-clearvr-drives-and-leverages-standards/}
\BIBentrySTDinterwordspacing

\bibitem{10071541}
N.~Sehad, B.~Cherif, I.~Khadraoui, W.~Hamidouche, F.~Bader, R.~Jäntti, and
  M.~Debbah, ``Locomotion-based {UAV} control toward the internet of senses,''
  \emph{IEEE Transactions on Circuits and Systems II: Express Briefs}, vol.~70,
  no.~5, pp. 1804--1808, 2023.

\bibitem{10.1145/3394171.3413999}
S.~Wang, X.~Zhang, M.~Xiao, K.~Chiu, and Y.~Liu, ``Spheric{RTC}: A system for
  content-adaptive real-time 360-degree video communication,'' in
  \emph{Proceedings of the 28th ACM International Conference on Multimedia},
  ser. MM '20.\hskip 1em plus 0.5em minus 0.4em\relax New York, NY, USA:
  Association for Computing Machinery, 2020, p. 3595–3603.

\bibitem{Intel360}
\BIBentryALTinterwordspacing
Intel. (2022, Dec.) Immersive video sample reference implementation. Accessed
  on 1-July-2023. [Online]. Available:
  \url{https://www.intel.com/content/www/us/en/developer/articles/technical/immersive-video-sample-powered-by-ovc.html}
\BIBentrySTDinterwordspacing

\bibitem{9830046}
Q.~Zhang, J.~Wei, S.~Wang, S.~Ma, and W.~Gao, ``Real{VR}: Efficient,
  economical, and quality-of-experience-driven {VR} video system based on
  {MPEG} {OMAF},'' \emph{IEEE Transactions on Multimedia}, pp. 1--15, 2022.

\bibitem{bonnineau2022perceptual}
C.~Bonnineau, W.~Hamidouche, J.~Fournier, N.~Sidaty, J.-F. Travers, and
  O.~D{\'e}forges, ``Perceptual quality assessment of {HEVC} and {VVC}
  standards for {8K} video,'' \emph{IEEE Trans. Broadcast.}, vol.~68, no.~1,
  pp. 246--253, Mar. 2022.

\bibitem{10.1145/3083187.3083190}
C.~Zhou, Z.~Li, and Y.~Liu, ``A measurement study of oculus 360 degree video
  streaming,'' in \emph{Proceedings of the 8th ACM on Multimedia Systems
  Conference}, ser. MMSys'17.\hskip 1em plus 0.5em minus 0.4em\relax New York,
  NY, USA: Association for Computing Machinery, 2017, p. 27–37.

\bibitem{VRIndForum}
\BIBentryALTinterwordspacing
V.~I. Forum. (2021, Jan.) {VR Industry Forum Guidelines V2.3}. Accessed on
  1-July-2023. [Online]. Available:
  \url{https://www.vr-if.org/wp-content/uploads/vrif2020.180.00-Guidelines-2.3_clean..pdf}
\BIBentrySTDinterwordspacing

\bibitem{10.11452980055.2980056}
F.~Qian, L.~Ji, B.~Han, and V.~Gopalakrishnan, ``Optimizing 360 video delivery
  over cellular networks,'' in \emph{Proc. 5th Workshop All Things Cellular:
  Oper., Appl. Challenges}, Oct. 2016, pp. 1--6.

\bibitem{TileMedia}
\BIBentryALTinterwordspacing
TileMedia. (2022, Jan.) {Tiled Media High Quality VR streaming}. Accessed on
  1-July-2023. [Online]. Available:
  \url{https://www.tiledmedia.com/high-quality-vr-streaming/}
\BIBentrySTDinterwordspacing

\bibitem{Intel_360_Implementation}
\BIBentryALTinterwordspacing
{W. Cheung, Gang Shen and Dusty Robbins}, ``Solution implementation summary:
  Media 360-degree video distribution,'' accessed: 19-january-2023. [Online].
  Available:
  \url{https://www.intel.it/content/dam/www/central-libraries/us/en/documents/advanced-360video-implementation-summary-final.pdf}
\BIBentrySTDinterwordspacing

\bibitem{Warburton2022}
M.~Warburton, M.~Mon-Williams, F.~Mushtaq, and J.~R. Morehead, ``Measuring
  motion-to-photon latency for sensorimotor experiments with virtual reality
  systems,'' \emph{Springer - Behavior Research Methods}, vol.~10, Oct. 2022.

\bibitem{7993736}
H.~T.~T. Tran, N.~P. Ngoc, C.~M. Bui, M.~H. Pham, and T.~C. Thang, ``An
  evaluation of quality metrics for 360 videos,'' in \emph{2017 Ninth
  International Conference on Ubiquitous and Future Networks (ICUFN)}, 2017,
  pp. 7--11.

\bibitem{GaoZenWanWuZhaSonQiaJin21}
N.~Gao, Y.~Zeng, J.~Wang, D.~Wu, C.~Zhang, Q.~Song, J.~Qian, and S.~Jin,
  ``Energy model for {UAV} communications: Experimental validation and model
  generalization,'' \emph{China Commun.}, vol.~18, no.~7, pp. 253--264, Jul.
  2021.

\bibitem{YanYanDinChe22}
H.~Yan, S.-H. Yang, Y.~Ding, and Y.~Chen, ``Energy consumption models for {UAV}
  communications: A brief survey,'' in \emph{Proc. IEEE Int. Conf. Internet
  Things and IEEE Green Comput. Commun. and IEEE Cyber, Physical Social Comput.
  and IEEE Smart Data and IEEE Congr. Cybermatics}, Aug. 2022, pp. 161--167.

\bibitem{You_TWC_Jun2018}
C.~You and R.~Zhang, ``{3D} trajectory optimization in {R}ician fading for
  {UAV}-enabled data harvesting,'' \emph{IEEE Transactions on Wireless
  Communications}, vol.~18, no.~6, pp. 3192--3207, 2019.

\bibitem{YouZha20}
------, ``Hybrid offline-online design for {UAV}-enabled data harvesting in
  probabilistic {LoS} channels,'' \emph{IEEE Trans. Wireless Commun.}, vol.~19,
  no.~6, pp. 3753--3768, Jun. 2020.

\bibitem{HouKanLar14}
A.~Al-Hourani, S.~Kandeepan, and S.~Lardner, ``Optimal {LAP} altitude for
  maximum coverage,'' \emph{IEEE Wireless Commun. Lett.}, vol.~3, no.~6, pp.
  569--572, Dec. 2014.

\bibitem{XiaWanCheCaoJiaZha20}
X.~Xiao, W.~Wang, T.~Chen, Y.~Cao, T.~Jiang, and Q.~Zhang, ``Sensor-augmented
  neural adaptive bitrate video streaming on {UAV}s,'' \emph{IEEE Trans.
  Multimedia}, vol.~22, no.~6, pp. 1567--1576, Jun. 2020.

\bibitem{MuzYanRafBetCav20}
R.~Muzaffar, E.~Yanmaz, C.~Raffelsberger, C.~Bettstetter, and A.~Cavallaro,
  ``Live multicast video streaming from drones: An experimental study,''
  \emph{Auton. Robot.}, vol.~44, pp. 75--91, 2020.

\bibitem{Cha19}
J.~Chakareski, ``{UAV-IoT} for next generation virtual reality,'' \emph{IEEE
  Trans. Image Process.}, vol.~28, no.~12, pp. 5977--5990, Dec. 2019.

\bibitem{HeXieTia19}
C.~He, Z.~Xie, and C.~Tian, ``A {QoE}-oriented uplink allocation for
  multi-{UAV} video streaming,'' \emph{Sensors}, vol.~19, no.~15, pp. 1--19,
  Aug. 2019.

\bibitem{ZhaMiaZhaYuFuWu20}
Q.~Zhang, J.~Miao, Z.~Zhang, F.~R. Yu, F.~Fu, and T.~Wu, ``Energy-efficient
  video streaming in {UAV}-enabled wireless networks: A safe-{DQN} approach,''
  in \emph{Proc. IEEE Global Commun. Conf.}, Dec. 2020, pp. 1--7.

\bibitem{ZhaCha22}
L.~Zhang and J.~Chakareski, ``{UAV}-assisted edge computing and streaming for
  wireless virtual reality: Analysis, algorithm design, and performance
  guarantees,'' \emph{IEEE Trans. Veh. Technol.}, vol.~71, no.~3, pp.
  3267--3275, Mar. 2022.

\bibitem{GuoCheHuZhe20}
Y.~Guo, Y.~Chen, J.~Hu, and H.~Zheng, ``Trajectory planning of {UAV} with
  real-time video transmission based on genetic algorithm,'' in \emph{Proc.
  Cross Strait Radio Sci. Wireless Technol. Conf.}, Dec. 2020, pp. 1--3.

\bibitem{ZhaHuWanFanNiy20}
C.~Zhan, H.~Hu, Z.~Wang, R.~Fan, and D.~Niyato, ``Unmanned aircraft system
  aided adaptive video streaming: A joint optimization approach,'' \emph{IEEE
  Trans. Multimedia}, vol.~22, no.~3, pp. 795--807, Mar. 2020.

\bibitem{BurLiuDenChaZah22}
L.~A.~b. Burhanuddin, X.~Liu, Y.~Deng, U.~Challita, and A.~Zahemszky, ``{QoE}
  optimization for live video streaming in {UAV-to-UAV} communications via deep
  reinforcement learning,'' \emph{IEEE Trans. Veh. Technol.}, vol.~71, no.~5,
  pp. 5358--5370, May 2022.

\bibitem{KhaChaGup20}
M.~Khan, J.~Chakareski, and S.~Gupta, ``{RF-FSO} dual-path {UAV} network for
  high fidelity multi-viewpoint scalable 360$^\circ$ video streaming,'' in
  \emph{Proc. IEEE 22nd Int. Workshop Multimedia Signal Process.}, Sep. 2020,
  pp. 1--6.

\bibitem{StoFakHelPopBet21}
A.~Stornig, A.~Fakhreddine, H.~Hellwagner, P.~Popovski, and C.~Bettstetter,
  ``Video quality and latency for {UAV} teleoperation over {LTE}: A study with
  ns3,'' in \emph{Proc. IEEE 93rd Veh. Technol. Conf.}, Apr. 2021, pp. 1--7.

\bibitem{ZhoHuJurMehDen21}
H.~Zhou, F.~Hu, M.~Juras, A.~B. Mehta, and Y.~Deng, ``Real-time video streaming
  and control of cellular-connected {UAV} system: Prototype and performance
  evaluation,'' \emph{IEEE Wireless Commun. Lett.}, vol.~10, no.~8, pp.
  1657--1661, Aug. 2021.

\bibitem{JinMaLiuLuWuHuaQin21}
J.~Jin, J.~Ma, L.~Liu, L.~Lu, G.~Wu, D.~Huang, and N.~Qin, ``Design of {UAV}
  video and control signal real-time transmission system based on {5G}
  network,'' in \emph{Proc. IEEE 16th Conf. Ind. Electron. Appl.}, Aug. 2021,
  pp. 533--537.

\bibitem{9951155}
T.~Taleb, N.~Sehad, Z.~Nadir, and J.~Song, ``{VR}-based immersive service
  management in b5g mobile systems: A uav command and control use case,''
  \emph{IEEE Internet of Things Journal}, vol.~10, no.~6, pp. 5349--5363, 2023.

\bibitem{QazSidWaq15}
S.~Qazi, A.~S. Siddiqui, and A.~I. Wagan, ``{UAV} based real time video
  surveillance over {4G LTE},'' in \emph{Proc. Int. Conf. Open Source Syst.
  Technol.}, Dec. 2015, pp. 141--145.

\bibitem{NavQazKhaMus19}
M.~Naveed, S.~Qazi, B.~A. Khawaja, and M.~Mustaqim, ``Evaluation of video
  streaming capacity of {UAV}s with respect to channel variation in {4G-LTE}
  surveillance architecture,'' in \emph{Proc. 8th Int. Conf. Inf. Commun.
  Technol.}, Nov. 2019, pp. 149--154.

\bibitem{SinCha21}
C.~Singhal and B.~N. Chandana, ``Aerial-{SON}: {UAV}-based self-organizing
  network for video streaming in dense urban scenario,'' in \emph{Proc. Int.
  Conf. Commun. Syst. Netw.}, Jan. 2021, pp. 7--12.

\bibitem{LiuJia22}
Z.~Liu and Y.~Jiang, ``Cross-layer design for {UAV}-based streaming media
  transmission,'' \emph{IEEE Trans. Circuits Syst. Video Technol.}, vol.~32,
  no.~7, pp. 4710--4723, Jul. 2022.

\bibitem{Nihei22}
K.~Nihei, N.~Kai, Y.~Maruyama, T.~Yamashita, D.~Kanetomo, T.~Kitahara,
  M.~Maruyama, T.~Ohki, K.~Kusin, and H.~Segah, ``Forest fire surveillance
  using live video streaming from {UAV} via multiple {LTE} networks,'' in
  \emph{Proc. IEEE 19th Annu. Consum. Commun. Netw. Conf.}, Jan. 2022, pp.
  465--468.

\bibitem{YuTakKaiSak21}
T.~Yu, Y.~Takaku, Y.~Kaieda, and K.~Sakaguchi, ``Design and {PoC}
  implementation of {mmWave}-based offloading-enabled {UAV} surveillance
  system,'' \emph{IEEE Open J. Veh. Technol.}, vol.~2, pp. 436--447, 2021.

\bibitem{HuDenAgh22}
F.~Hu, Y.~Deng, and A.~H. Aghvami, ``Cooperative multigroup broadcast 360°
  video delivery network: A hierarchical federated deep reinforcement learning
  approach,'' \emph{IEEE Trans. Wireless Commun.}, vol.~21, no.~6, pp.
  4009--4024, Jun. 2022.

\bibitem{Lin18}
X.~Lin, V.~Yajnanarayana, S.~D. Muruganathan, S.~Gao, H.~Asplund, H.-L.
  Määttänen, M.~Bergstrom, S.~Euler, and Y.-P.~E. Wang, ``The sky is not the
  limit: {LTE} for unmanned aerial vehicles,'' \emph{IEEE Commun. Mag.},
  vol.~56, no.~4, pp. 204--210, Apr. 2018.

\bibitem{3GPP_TR_36777}
\BIBentryALTinterwordspacing
{3GPP TR 36.777}, ``Enhanced {LTE} support for aerial vehicles.'' [Online].
  Available:
  \url{https://portal.3gpp.org/desktopmodules/Specifications/SpecificationDetails.aspx?specificationId=3231}
\BIBentrySTDinterwordspacing

\bibitem{3GPP_TS_22125_Rel16}
\BIBentryALTinterwordspacing
{3GPP TS 22.125}, ``Unmanned aerial system ({UAS}) support in {3GPP},''
  accessed: 1-July-2023. [Online]. Available:
  \url{https://portal.3gpp.org/desktopmodules/Specifications/SpecificationDetails.aspx?specificationId=3545}
\BIBentrySTDinterwordspacing

\bibitem{3GPP_ATIS_Rel17_UAV}
\BIBentryALTinterwordspacing
{ATIS-I-0000092}, ``{3GPP Release} 17 – building blocks for uav
  applications,'' accessed: 1-July-2023. [Online]. Available:
  \url{https://access.atis.org/apps/group_public/download.php/66824/ATIS-I-0000092.pdf}
\BIBentrySTDinterwordspacing

\bibitem{3GPP_TR_26918_Rel15}
\BIBentryALTinterwordspacing
{3GPP TS 26.918}, ``Virtual reality ({VR}) media services over 3gpp,''
  accessed: 1-July-2023. [Online]. Available:
  \url{https://portal.3gpp.org/desktopmodules/Specifications/SpecificationDetails.aspx?specificationId=3053}
\BIBentrySTDinterwordspacing

\bibitem{3GPP_TS_26501_Rel18}
\BIBentryALTinterwordspacing
{3GPP TS 26.501}, ``{5G} media streaming ({5GMS}); general description and
  architecture,'' accessed: 1-July-2023. [Online]. Available:
  \url{https://portal.3gpp.org/desktopmodules/Specifications/SpecificationDetails.aspx?specificationId=3582}
\BIBentrySTDinterwordspacing

\bibitem{3GPP_TS_26506_Rel18}
\BIBentryALTinterwordspacing
{3GPP TS 26.506}, ``{5G} real-time media communication architecture,''
  accessed: 1-July-2023. [Online]. Available:
  \url{https://portal.3gpp.org/desktopmodules/Specifications/SpecificationDetails.aspx?specificationId=4102}
\BIBentrySTDinterwordspacing

\bibitem{3GPP_TS_26522_Rel18}
\BIBentryALTinterwordspacing
{3GPP TS 26.522}, ``{5G} real-time media transport protocol configurations,''
  accessed: 1-July-2023. [Online]. Available:
  \url{https://portal.3gpp.org/desktopmodules/Specifications/SpecificationDetails.aspx?specificationId=4114}
\BIBentrySTDinterwordspacing

\bibitem{3GPP_TS_26803_Rel18}
\BIBentryALTinterwordspacing
{3GPP TS 26.803}, ``Study on {5G} media streaming extensions for edge
  processing,'' accessed: 1-July-2023. [Online]. Available:
  \url{https://portal.3gpp.org/desktopmodules/Specifications/SpecificationDetails.aspx?specificationId=3742}
\BIBentrySTDinterwordspacing

\bibitem{3GPP_TR_26927_Rel18}
\BIBentryALTinterwordspacing
{\relax 3GPP TR 26.927}, ``{Study on Artificial Intelligence and Machine
  learning in {5G} media services},'' accessed: 1-July-2023. [Online].
  Available:
  \url{https://portal.3gpp.org/desktopmodules/Specifications/SpecificationDetails.aspx?specificationId=4040}
\BIBentrySTDinterwordspacing

\bibitem{Perrin_MDPI_2020}
A.-F. Perrin, V.~Krassanakis, L.~Zhang, V.~Ricordel, M.~Perreira Da~Silva, and
  O.~Le~Meur, ``{EyeTrackUAV2}: A large-scale binocular eye-tracking dataset
  for {UAV} videos,'' \emph{Drones}, vol.~4, no.~1, 2020.

\bibitem{Fu_TIP_2020}
K.~Fu, J.~Li, Y.~Zhang, H.~Shen, and Y.~Tian, ``Model-guided multi-path
  knowledge aggregation for aerial saliency prediction,'' \emph{IEEE Trans.
  Image Process.}, vol.~29, pp. 7117--7127, 2020.

\bibitem{Colonnese_EUVIP_2018}
S.~Colonnese, F.~Cuomo, L.~Ferranti, and T.~Melodia, ``Efficient video
  streaming of 360° cameras in unmanned aerial vehicles: An analysis of real
  video sources,'' in \emph{Proc. 7th Eur. Workshop Visual Inf. Process.}, Nov.
  2018, pp. 1--6.

\bibitem{Mi_applied_sciences_2019}
T.-W. Mi and M.-T. Yang, ``Comparison of tracking techniques on 360-degree
  videos,'' \emph{Appl. Sci.}, vol.~9, no.~16, p. 3336, 2019.

\bibitem{David_MMSys_2018}
E.~J. David, J.~Guti\'{e}rrez, A.~Coutrot, M.~P. Da~Silva, and P.~L. Callet,
  ``A dataset of head and eye movements for 360° videos,'' in \emph{Proc. 9th
  ACM Multimedia Syst. Conf.}, Jun. 2018, pp. 432--437.

\bibitem{Nasrabadi_MultiSys_conf_2019}
A.~T. Nasrabadi, A.~Samiei, A.~Mahzari, R.~P. McMahan, R.~Prakash, M.~C.~Q.
  Farias, and M.~M. Carvalho, ``A taxonomy and dataset for 360° videos,'' in
  \emph{Proc. 10th ACM Multimedia Syst. Conf.}, Jun. 2019, pp. 273--278.

\bibitem{Wu_MMSys_2017}
C.~Wu, Z.~Tan, Z.~Wang, and S.~Yang, ``A dataset for exploring user behaviors
  in {VR} spherical video streaming,'' in \emph{Proc. 8th ACM Multimedia Syst.
  Conf.}, Jun. 2017, pp. 193--198.

\bibitem{libx264}
{VideoLAN}, ``libx264,'' \url{https://www.videolan.org/developers/x264.html},
  2013, accessed: 1-July-2023.

\bibitem{h264_nvenc}
{NVIDIA Corporation}, ``h264\_nvenc,''
  \url{https://developer.nvidia.com/nvidia-video-codec-sdk}, 2018, accessed:
  1-July-2023.

\bibitem{x265}
{MulticoreWare Inc.}, ``x265,''
  \url{https://www.videolan.org/developers/x265.html}, 2021, version: 3.5.

\bibitem{hevc_nvenc}
{NVIDIA Corporation}, ``hevc\_nvenc,''
  \url{https://developer.nvidia.com/nvidia-video-codec-sdk}, May 2021, sDK
  Version: 11.0.

\bibitem{9455944}
A.~Wieckowski, J.~Brandenburg, T.~Hinz, C.~Bartnik, V.~George, G.~Hege,
  C.~Helmrich, A.~Henkel, C.~Lehmann, C.~Stoffers, I.~Zupancic, B.~Bross, and
  D.~Marpe, ``Vvenc: An open and optimized vvc encoder implementation,'' in
  \emph{2021 IEEE International Conference on Multimedia \& Expo Workshops
  (ICMEW)}, 2021, pp. 1--2.

\bibitem{libsvt_av1}
{Open Visual Cloud}, ``Svt-av1,'' \url{https://www.openvisualcloud.org/}, 2021,
  version: 0.8.7.

\bibitem{libvpx_vp9}
{WebM Project}, ``libvpx,'' \url{https://www.webmproject.org/code/}, 2021,
  version: 1.10.0.

\bibitem{huang2021deep}
T.~Huang, R.-X. Zhang, and L.~Sun, ``Deep reinforced bitrate ladders for
  adaptive video streaming,'' in \emph{Proceedings of the 31st ACM Workshop on
  Network and Operating Systems Support for Digital Audio and Video}, 2021, pp.
  66--73.

\bibitem{mandhane2022muzero}
A.~Mandhane, A.~Zhernov, M.~Rauh, C.~Gu, M.~Wang, F.~Xue, W.~Shang, D.~Pang,
  R.~Claus, C.-H. Chiang \emph{et~al.}, ``Muzero with self-competition for rate
  control in vp9 video compression,'' \emph{arXiv preprint arXiv:2202.06626},
  2022.

\bibitem{IntelTrueview}
\BIBentryALTinterwordspacing
Intel, ``True view intel sports,'' 2021, accessed: 1-July-2023. [Online].
  Available:
  \url{https://www.intel.com/content/www/us/en/sports/technology/true-view.html}
\BIBentrySTDinterwordspacing

\bibitem{Wang_ICDCS_2017}
X.~Wang, A.~Chowdhery, and M.~Chiang, ``Networked drone cameras for sports
  streaming,'' in \emph{Proc. IEEE 37th Int. Conf. Distrib. Comput. Syst.},
  Jun. 2017, pp. 308--318.

\bibitem{Meng_ACMMulti_2015}
X.~Meng, W.~Wang, and B.~Leong, ``{SkyStitch}: A cooperative multi-{UAV}-based
  real-time video surveillance system with stitching,'' in \emph{Proc. 23rd ACM
  Int. Conf. Multimedia}, Oct. 2015, pp. 261--270.

\bibitem{Wang_MILCOM2022}
Y.~Wang and J.~Farooq, ``Zero touch coordinated uav network formation for
  $360^{\circ}$ views of a moving ground target in remote {VR} applications,''
  in \emph{Proc. IEEE Military Commun. Conf.}, Nov. 2022, pp. 950--955.

\bibitem{WanDuLiSon21}
Z.~Wang, J.~Du, G.~Li, and X.~Song, ``Control of panoramic video flow for
  {UAV},'' in \emph{Proc. Int. Conf. Social Comput. Digit. Economy}, Aug. 2021,
  pp. 105--107.

\bibitem{QiaWanJin22}
J.~Qian, J.~Wang, and S.~Jin, ``Configurable virtual {MIMO} via {UAV} swarm:
  Channel modeling and spatial correlation analysis,'' \emph{China Commun.},
  vol.~19, no.~9, pp. 133--145, Sep. 2022.

\bibitem{Anwar_IEEEAccess_2020}
M.~S. Anwar, J.~Wang, W.~Khan, A.~Ullah, S.~Ahmad, and Z.~Fei, ``Subjective
  {QoE} of 360-degree virtual reality videos and machine learning
  predictions,'' \emph{IEEE Access}, vol.~8, pp. 148\,084--148\,099, 2020.

\bibitem{Teng_TWC_Jan2022}
L.~Teng, G.~Zhai, Y.~Wu, X.~Min, W.~Zhang, Z.~Ding, and C.~Xiao, ``{QoE} driven
  {VR} 360° video massive {MIMO} transmission,'' \emph{IEEE Trans. Wireless
  Commun.}, vol.~21, no.~1, pp. 18--33, Jan. 2022.

\bibitem{Huang_ArXiv2018}
W.~Huang, L.~Ding, G.~Zhai, X.~Min, J.-N. Hwang, Y.~Xu, and W.~Zhang,
  ``Utility-oriented resource allocation for 360-degree video transmission over
  heterogeneous networks,'' \emph{Digit. Signal Process.}, vol.~84, pp. 1--14,
  Jan. 2019.

\bibitem{Cristina_TC_Apr2020}
C.~Perfecto, M.~S. Elbamby, J.~D. Ser, and M.~Bennis, ``Taming the latency in
  multi-user {VR} 360°: A {QoE}-aware deep learning-aided multicast
  framework,'' \emph{IEEE Trans. Commun.}, vol.~68, no.~4, pp. 2491--2508, Apr.
  2020.

\bibitem{Ding_Globecom_2019}
Y.~Ding, D.~Jiang, J.~Huang, L.~Xiao, S.~Liu, Y.~Tang, and H.~Dai,
  ``{QoE}-aware power control for {UAV}-aided media transmission with
  reinforcement learning,'' in \emph{Proc. IEEE Global Commun. Conf.}, Dec.
  2019, pp. 1--6.

\bibitem{Shirey_ISQoS_2021}
R.~Shirey, S.~Rao, and S.~Sundaram, ``Optimizing quality of experience for
  long-range {UAS} video streaming,'' in \emph{Proc. IEEE/ACM 29th Int. Symp.
  Quality Service}, Jun. 2021, pp. 1--10.

\bibitem{Colonnese_IEEESAM_2018}
S.~Colonnese, A.~Carlesimo, L.~Brigato, and F.~Cuomo, ``{QoE}-aware {UAV}
  flight path design for mobile video streaming in {HetNet},'' in \emph{Proc.
  IEEE 10th Sensor Array Multichannel Signal Process. Workshop}, Jul. 2018, pp.
  301--305.

\bibitem{Tang_TMC_July2022}
M.~Tang and V.~W. Wong, ``Online bitrate selection for viewport adaptive
  360-degree video streaming,'' \emph{IEEE Trans. Mobile Comput.}, vol.~21,
  no.~7, pp. 2506--2517, Jul. 2022.

\bibitem{Yang_IEEEAccess_2019}
J.~Yang, J.~Luo, D.~Meng, and J.-N. Hwang, ``{QoE}-driven resource allocation
  optimized for uplink delivery of delay-sensitive {VR} video over cellular
  network,'' \emph{IEEE Access}, vol.~7, pp. 60\,672--60\,683, 2019.

\bibitem{Liyana_TVT_May2022}
L.~A.~b. Burhanuddin, X.~Liu, Y.~Deng, U.~Challita, and A.~Zahemszky, ``{QoE}
  optimization for live video streaming in {UAV-to-UAV} communications via deep
  reinforcement learning,'' \emph{IEEE Trans. Veh. Technol.}, vol.~71, no.~5,
  pp. 5358--5370, May 2022.

\bibitem{LiuZhuDenGuaWanLuoLinZha19}
Y.~Liu, C.~Zhu, X.~Deng, P.~Guan, Z.~Wan, J.~Luo, E.~Liu, and H.~Zhang,
  ``{UAV}-aided urban target tracking system based on edge computing,''
  \emph{CoRR}, vol. abs/1902.00837, pp. 1--6, Feb. 2019.

\bibitem{CalBar21}
E.~{Calvanese Strinati} and S.~Barbarossa, ``{6G} networks: Beyond {Shannon}
  towards semantic and goal-oriented communications,'' \emph{Computer
  Networks}, vol. 190, p. 107930, May 2021.

\bibitem{47751}
\BIBentryALTinterwordspacing
J.~Devlin, M.-W. Chang, K.~Lee, and K.~N. Toutanova, ``{BERT}: Pre-training of
  deep bidirectional transformers for language understanding,'' 2018. [Online].
  Available: \url{https://arxiv.org/abs/1810.04805}
\BIBentrySTDinterwordspacing

\bibitem{bariah2023large}
L.~Bariah, Q.~Zhao, H.~Zou, Y.~Tian, F.~Bader, and M.~Debbah, ``Large language
  models for telecom: The next big thing?'' \emph{arXiv preprint
  arXiv:2306.10249}, 2023.

\bibitem{vemprala2023chatgpt}
\BIBentryALTinterwordspacing
S.~Vemprala, R.~Bonatti, A.~Bucker, and A.~Kapoor, ``Chat{GPT} for robotics:
  Design principles and model abilities,'' Microsoft, Tech. Rep. MSR-TR-2023-8,
  February 2023. [Online]. Available:
  \url{https://www.microsoft.com/en-us/research/publication/chatgpt-for-robotics-design-principles-and-model-abilities/}
\BIBentrySTDinterwordspacing

\bibitem{wang2022simvlm}
\BIBentryALTinterwordspacing
Z.~Wang, J.~Yu, A.~W. Yu, Z.~Dai, Y.~Tsvetkov, and Y.~Cao, ``Sim{VLM}: Simple
  visual language model pretraining with weak supervision,'' in
  \emph{International Conference on Learning Representations}, 2022. [Online].
  Available: \url{https://openreview.net/forum?id=GUrhfTuf_3}
\BIBentrySTDinterwordspacing

\bibitem{alayrac2022flamingo}
\BIBentryALTinterwordspacing
J.-B. Alayrac, J.~Donahue, P.~Luc, A.~Miech, I.~Barr, Y.~Hasson, K.~Lenc,
  A.~Mensch, K.~Millican, M.~Reynolds, R.~Ring, E.~Rutherford, S.~Cabi, T.~Han,
  Z.~Gong, S.~Samangooei, M.~Monteiro, J.~Menick, S.~Borgeaud, A.~Brock,
  A.~Nematzadeh, S.~Sharifzadeh, M.~Binkowski, R.~Barreira, O.~Vinyals,
  A.~Zisserman, and K.~Simonyan, ``Flamingo: a visual language model for
  few-shot learning,'' in \emph{Advances in Neural Information Processing
  Systems}, A.~H. Oh, A.~Agarwal, D.~Belgrave, and K.~Cho, Eds., 2022.
  [Online]. Available: \url{https://openreview.net/forum?id=EbMuimAbPbs}
\BIBentrySTDinterwordspacing

\bibitem{li2023blip2}
J.~Li, D.~Li, S.~Savarese, and S.~Hoi, ``Blip-2: Bootstrapping language-image
  pre-training with frozen image encoders and large language models,'' 2023.

\end{thebibliography}

\end{document}